\documentclass[a4paper, 12pt]{article}\usepackage[]{graphicx}\usepackage[dvipsnames,table]{xcolor}
\makeatletter
\def\maxwidth{ %
  \ifdim\Gin@nat@width>\linewidth
    \linewidth
  \else
    \Gin@nat@width
  \fi
}
\makeatother

\definecolor{fgcolor}{rgb}{0.345, 0.345, 0.345}
\newcommand{\hlnum}[1]{\textcolor[rgb]{0.686,0.059,0.569}{#1}}%
\newcommand{\hlsng}[1]{\textcolor[rgb]{0.192,0.494,0.8}{#1}}%
\newcommand{\hlcom}[1]{\textcolor[rgb]{0.678,0.584,0.686}{\textit{#1}}}%
\newcommand{\hlopt}[1]{\textcolor[rgb]{0,0,0}{#1}}%
\newcommand{\hldef}[1]{\textcolor[rgb]{0.345,0.345,0.345}{#1}}%
\newcommand{\hlkwb}[1]{\textcolor[rgb]{0.69,0.353,0.396}{#1}}%
\newcommand{\hlkwc}[1]{\textcolor[rgb]{0.333,0.667,0.333}{#1}}%
\newcommand{\hlkwd}[1]{\textcolor[rgb]{0.737,0.353,0.396}{\textbf{#1}}}%

\usepackage{framed}
\makeatletter
\newenvironment{kframe}{%
 \def\at@end@of@kframe{}%
 \ifinner\ifhmode%
  \def\at@end@of@kframe{\end{minipage}}%
  \begin{minipage}{\columnwidth}%
 \fi\fi%
 \def\FrameCommand##1{\hskip\@totalleftmargin \hskip-\fboxsep
 \colorbox{shadecolor}{##1}\hskip-\fboxsep
     \hskip-\linewidth \hskip-\@totalleftmargin \hskip\columnwidth}%
 \MakeFramed {\advance\hsize-\width
   \@totalleftmargin\z@ \linewidth\hsize
   \@setminipage}}%
 {\par\unskip\endMakeFramed%
 \at@end@of@kframe}
\makeatother

\definecolor{shadecolor}{rgb}{.97, .97, .97}
\definecolor{messagecolor}{rgb}{0, 0, 0}
\definecolor{warningcolor}{rgb}{1, 0, 1}
\definecolor{errorcolor}{rgb}{1, 0, 0}
\newenvironment{knitrout}{}{} 

\usepackage{alltt}
\usepackage{amsmath, amssymb} 
\usepackage{array} 
\usepackage{doi} 
\usepackage[numbers,sort&compress]{natbib}
\usepackage{multirow} 
\usepackage{booktabs} 
\usepackage{longtable}
\usepackage[title]{appendix} 
\usepackage{nameref} 
\usepackage[dvipsnames,table]{xcolor}
\usepackage[doublespacing]{setspace} 
\usepackage{helvet}
\usepackage{mathpazo}
\usepackage{sectsty} 
\allsectionsfont{\sffamily} 
\usepackage[labelfont={bf,sf},font=small]{caption} 
\usepackage{orcidlink} 
\usepackage{todonotes}
\usepackage{tcolorbox}
\tcbuselibrary{breakable}

\usepackage{geometry}
\title{\vspace{-4em} \textbf{\textsf{
      Edgington's Combination Method for Two-Study Meta-Analysis: An Empirical Evaluation in 1226 Meta-Analyses
}}}

\usepackage{authblk}

\author[1]{Samuel~Pawel~\orcidlink{0000-0003-2779-320X}}
\author[1]{Saverio~Fontana~\orcidlink{0009-0001-9116-7812}}
\author[1,2]{Jinyu~Chen~\orcidlink{0009-0005-6803-9555}}
\author[3]{Leonie~Stoltefuß~\orcidlink{0000-0001-9811-3641}}
\author[3]{Frank~Weber~\orcidlink{0000-0002-4842-7922}}
\author[3]{Guido~Skipka}
\author[3]{\authorcr Sibylle~Sturtz}
\author[3]{Ralf~Bender~\orcidlink{0000-0002-2422-4362}}
\author[1]{Leonhard~Held~\orcidlink{0000-0002-8686-5325}}

\affil[1]{Epidemiology, Biostatistics and Prevention Institute (EBPI),
  Center for Reproducible Science and Research Synthesis (CRS),
  University of Zurich,
  Zürich, Switzerland}
\affil[2]{Seminar for Statistics,
  ETH Zurich,
  Zürich, Switzerland}
\affil[3]{Department of Medical Biometry,
  Institute for Quality and Efficiency in Health Care (IQWiG),
  Köln, Germany}
\date{July 27, 2026}

\usepackage{hyperref}
\hypersetup{
  bookmarksopen=true,
  breaklinks=true,
  pdfsubject={},
  pdfkeywords={},
  colorlinks=true,
  linkcolor=blue,
  anchorcolor=black,
  citecolor=blue,
  urlcolor=blue,
}

\IfFileExists{upquote.sty}{\usepackage{upquote}}{}
\begin{document}

\begin{onehalfspacing}
\maketitle

\begin{abstract}
  \noindent Two-study meta-analyses are common in evidence synthesis but pose major statistical challenges. With only two studies, the between-study variance cannot be reliably estimated, rendering standard random-effects methods unstable. Here, we investigate meta-analyses based on Edgington's \textit{p}-value combination method as an alternative approach, applying it to 1226 two-study meta-analyses from the German Institute for Quality and Efficiency in Health Care (IQWiG). Like fixed-effect meta-analysis, Edgington's method is calibrated under homogeneity. However, it adapts confidence interval width to observed between-study discrepancy without requiring explicit heterogeneity estimation. In all of the examined meta-analyses, this leads to confidence intervals that contain both study-specific estimates but remain informative. Edgington's method agrees with fixed-effect meta-analysis on statistical significance (at two-sided $\alpha = 0.05$) in 91\% of all meta-analyses, but can give wider intervals when study results are discrepant and narrower intervals when results are highly consistent. Weighted extensions of Edgington's method shift point estimates toward the more precise study while preserving much of this adaptive behavior. We conclude that Edgington's method offers a principled and practically useful complement to existing approaches for two-study meta-analysis, occupying a middle ground between standard fixed-effect and random-effects approaches. 
\end{abstract}

\begin{tcolorbox}[
    breakable,
    title = {Highlights},
    fonttitle = \bfseries\sffamily\large,
    colback = white,
    colframe = black!70,
    colbacktitle = black!10,
    coltitle = black,
    boxrule = 0.75pt
]

\subsubsection*{What is already known}

\begin{itemize}
    \item Two-study meta-analyses are common in evidence synthesis, accounting
      for more than half of all meta-analyses in the IQWiG database.

    \item Standard random-effects methods lead to unstable inferences with only
      two studies, as between-study variance cannot be reliably estimated.

    \item Edgington's combination method is a valid fixed-effect meta-analysis
      approach whose confidence interval width adapts to between-study
      discrepancy, unlike standard fixed-effect methods.
\end{itemize}

\subsubsection*{What is new}

\begin{itemize}
    \item We evaluate unweighted and weighted Edgington's method on 1226
      two-study meta-analyses from IQWiG, comparing confidence intervals and
      point estimates to standard fixed-effect and random-effects methods.

    \item Edgington's confidence intervals contain both study-specific
      estimates, remain informative, and agree with fixed-effect meta-analysis
      on statistical significance in the majority of cases, while adapting to
      between-study discrepancy.

    \item Weighted extensions shift point estimates toward the more precise
      study while preserving much of the adaptive behavior.
\end{itemize}

\subsubsection*{Potential impact}

\begin{itemize}
    \item Edgington's method offers a principled middle ground between
      fixed-effect and random-effects meta-analysis for the two-study setting,
      and is freely available via the \texttt{confMeta} R package on CRAN.
\end{itemize}

\end{tcolorbox}

\end{onehalfspacing}

\clearpage

\section{Introduction}

Meta-analysis combines results from multiple studies to increase the precision
of effect estimates and provide a quantitative summary of the available
evidence. Typically, meta-analyses are conducted under either a fixed-effect
(also known as common-effect or equal-effect) or a random-effects model. The
fixed-effect model assumes that all studies estimate a common true effect, and
that differences between study results arise solely from sampling variability
within the studies. The random-effects model allows for heterogeneity in the
underlying true effect sizes across studies by assuming that they are sampled
from an overarching distribution, in which the variance models the degree of
heterogeneity between the studies \citep{Hedges1985, Borenstein2009, Egger2022}.

\begin{figure}[!htb]
\begin{knitrout}
\definecolor{shadecolor}{rgb}{0.969, 0.969, 0.969}\color{fgcolor}

{\centering \includegraphics[width=\maxwidth]{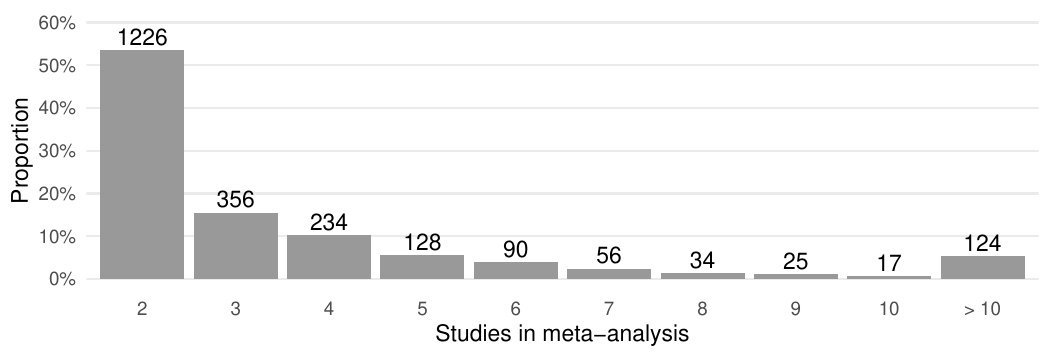} 

}

\end{knitrout}
\caption{Distribution of the number of studies per meta-analysis in IQWiG
  meta-analyses as of 2025.}
\label{fig:IQWiG-nstudies}
\end{figure}

While random-effects models formally account for heterogeneity, their
performance critically depends on reliable estimation of the between-study
variance. When only a small number of studies is available, estimation of this
variance becomes unstable. The most extreme case is a meta-analysis of only two
studies. These are surprisingly common. In an analysis of the Cochrane Database
of Systematic Reviews, more than one third (36\%) of all meta-analyses included
only two studies \citep{Davey2011}. Similarly, in the database which collects
meta-analyses from all published reports of the German Institute for Quality and
Efficiency in Health Care (IQWiG), more than 50\% of all meta-analyses contain
only two studies, see Figure~\ref{fig:IQWiG-nstudies}.

With only two observations, estimating a variance parameter is inherently
difficult. As a consequence, random-effects meta-analysis may yield highly
imprecise and unstable inference. Confidence intervals (CIs) can become
excessively wide and heterogeneity estimates highly variable \citep{Guolo2015,
  Bender2018, Schulz2021}. Therefore, developing methods that provide meaningful
quantitative inference in the two-study setting is of substantial importance and
remains an open problem \citep{Gonnermann2015, Bender2018}.

\begin{figure}[!htb]
\begin{knitrout}
\definecolor{shadecolor}{rgb}{0.969, 0.969, 0.969}\color{fgcolor}

{\centering \includegraphics[width=\maxwidth]{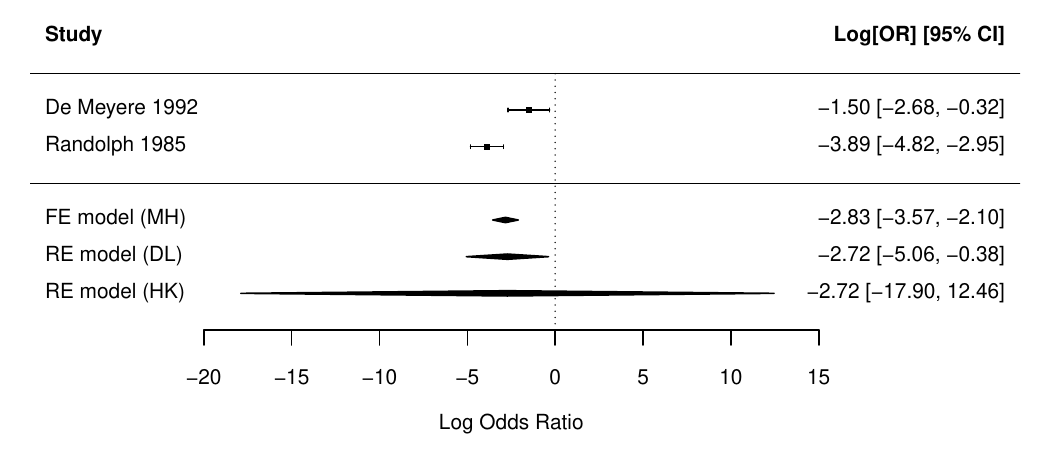} 

}

\end{knitrout}

\caption{Meta-analysis of the effect of antibiotics compared to placebo on sore
  throat \citep[Figure~2]{IQWiG2022}. The fixed-effect model is fitted using the
  Mantel-Haenszel (MH) approach (the standard IQWiG approach for meta-analysis
  with two studies and binary data). The random-effects models are fitted using
  the DerSimonian-Laird (DL) and the Hartung-Knapp (HK) approach with the
  Paule-Mandel method for estimating the heterogeneity ($\hat{\tau} =
  1.6$ with 95\% CI from
  $0.52$ to
  $10$).}
\label{fig:metaexample}
\end{figure}

Figure~\ref{fig:metaexample} illustrates this challenge with a two-study
meta-analysis on the effect of antibiotics compared to a placebo on sore throat
from the IQWiG database \citep{IQWiG2022}. For meta-analyses of at least five
studies, IQWiG uses the random-effects model with the Hartung-Knapp (HK)
adjustment and Paule-Mandel (PM) heterogeneity estimator\footnote{For two-study
meta-analyses, the PM heterogeneity estimator coincides with the
DerSimonian-Laird (DL) method of moments and the restricted maximum likelihood
(REML) estimators \citep[Section~5.5]{Jackson2017}. In this paper we will focus
on PM but all results apply also to DL and REML due to this equivalence.}. This
approach is recommended in the literature as the resulting CIs and
\textit{p}-values are calibrated in many scenarios \citep{Langan2018}. However,
for two-study meta-analyses, the HK approach is usually of limited practical use
as it leads to very wide CIs, which is also visible in
Figure~\ref{fig:metaexample}. Even though both study effect estimates and 95\%
CIs lie well below zero (indicating a treatment benefit), the HK interval is
uninformative, spanning an extremely large range from
$-17.90$ to $12.46$ (in terms of the log odds ratio) that extends well beyond the
range of the individual study CIs. The CI from $-5.06$ to $-0.38$ based on the standard DerSimonian-Laird (DL)
random-effects method (also known as as Wald-type \textit{z} method, see, e.g.,
Veroniki et al. \citep{Veroniki2016}) is less wide. However, it has been shown
that DL is unreliable for meta-analysis with few studies, often producing CIs
that are too narrow when there is true heterogeneity
\citep[e.g.,][]{Cornell2014, Langan2018}. IQWiG therefore follows a special
procedure in the two-study setting using fixed-effect meta-analysis by default,
and resorting to qualitative evidence synthesis when substantial heterogeneity
is observed \citep{Schulz2021, IQWiG2025}.

Several strategies have been proposed to address this challenge. One principled
approach is Bayesian meta-analysis \citep{Sutton2001}, which stabilizes
inference in the two-study setting by placing an informative prior distribution
on the heterogeneity parameter. Prior distributions are often chosen to be only
``weakly'' informative \citep{Roever2021} or derived from large collections of
meta-analyses \citep{Lilienthal2023}. Nevertheless, prior specification remains
the most challenging and controversial aspect of Bayesian meta-analysis,
especially with two studies, where the results depend strongly on the chosen
prior.

While Bayesian methods require prior specification, an alternative frequentist
approach based on \textit{p}-value combination avoids this need \citep{Xie2013,
  SchwederHjort2016}. In particular, Edgington's combination method
\citep{Edgington1972}, based on the sum of \textit{p}-values, has been shown to
have several useful properties for meta-analysis. Under homogeneity, it is
calibrated, like fixed-effect meta-analysis, but its CIs can widen with
increasing observed between-study discrepancy without requiring explicit
estimation of a heterogeneity parameter. A simulation study demonstrated that
this behavior translates into practical gains: Edgington's method improves upon
standard fixed-effect meta-analysis in settings with only three studies,
maintaining exact calibration under homogeneity and exhibiting less
miscalibration when true heterogeneity is present \citep{Held2025}. Theoretical
analysis of the two-study case showed that, as the individual study standard
errors decrease, the CI from Edgington's method always contains both true study
effects, and collapses to a point only when the two studies share a common true
effect \citep{Pawel2025}. Fixed-effect meta-analysis instead collapses to a
point at a weighted average of the true effects, even when the studies do not
share a common effect. Thus, the method asymptotically adapts to heterogeneity
in a different way to fixed-effect meta-analysis, yet it is unclear whether the
method provides any benefits in real data.

The aim of this paper is to evaluate Edgington's method as a practical
alternative to fixed-effect meta-analysis in the two-study setting. Our
evaluation is based on a large empirical database of 1226 two-study meta-analyses from IQWiG. A central question is
whether the method can improve upon IQWiG's current fixed-effect procedure in
the presence of observed heterogeneity. We also develop weighted extensions of
the method and assess whether these offer further improvements in empirical
data.

Section~\ref{sec:methods} describes Edgington's method, introduces weighted
extensions, and derives conditions under which Edgington's CIs are wider or
narrower than those of fixed-effect meta-analysis. Section~\ref{sec:iqwig}
applies these methods to the IQWiG database of two-study meta-analyses,
comparing them to fixed-effect and random-effects meta-analysis. The appendices
provide technical details, additional empirical results, and a demonstration of
how to perform a meta-analysis using Edgington's method with the
\texttt{confMeta} R package \citep{Hofmann2026}.

\section{Methods}
\label{sec:methods}
Suppose we want to synthesize the results of two studies in a meta-analysis.
Denote by $\hat{\theta}_i$ the estimate of the effect $\theta_i$ from study $i
\in \{1, 2\}$, and by $\sigma_i$ the corresponding standard error. Assuming
approximate normality, the associated one-sided \textit{p}-value for the null
hypothesis $H_{0i} \colon \theta_i = \mu$ and alternative $H_{1i} \colon
\theta_i > \mu$ is then $p_i(\mu) = 1 - \Phi\{(\hat{\theta}_i - \mu)/\sigma_i\}$
where $\Phi(\cdot)$ is the standard normal cumulative distribution function.
Viewing $p_i(\mu)$ as a function of the null value $\mu$ yields a so-called
\textit{p}-value function, also known as confidence distribution \citep[see,
  e.g.,][]{Bender2005, Xie2013, Infanger2019, Marschner2024}.

Various methods can be used to combine \textit{p}-value functions from different
studies into a meta-analytic \textit{p}-value function from which point
estimates and CIs can be obtained \citep{Xie2011}. Many of these depend on the
orientation of the study-specific \textit{p}-value functions, that is, a
different result is obtained if a flipped alternative hypothesis $H_{1i} \colon
\theta_i < \mu$ is specified. This is not the case for Edgington's method, which
is orientation-invariant, and which we consider in the following
\citep{Held2025}. Note that Edgington's method considered in this paper is not
taking into account a heterogeneity parameter, in contrast to the
heterogeneity-considering variants presented by Held et al. \citep{Held2025}.
Moreover, while Edgington's method can in principle combine any valid
\textit{p}-values, we will here use \textit{p}-values derived under normal
approximations.

\subsection{Edgington's method}

Edgington's method is based on the sum of the \textit{p}-values $s(\mu) =
p_1(\mu) + p_2(\mu)$. Under the null hypothesis, a \textit{p}-value has a
uniform distribution. Assuming that all study-specific null hypotheses are true,
the sum has an Irwin-Hall distribution with parameter $n = 2$ \citep{Irwin1927,
  Hall1927}, in this case a triangular distribution. A combined \textit{p}-value
can be computed from the corresponding cumulative distribution function by
\begin{align}
  \label{eq:pE}
  p_E(\mu) =
  \begin{cases}
     \{s(\mu)\}^2/2 & \text{if} ~ 0 \leq s(\mu) \leq 1 \\
     1 - \{2 - s(\mu)\}^2/2 & \text{if} ~ 1 < s(\mu) \leq 2. \\
  \end{cases}
\end{align}
Appendix~\ref{app:morethan2studies} shows the general form of Edgington's
combined \textit{p}-value for meta-analysis of any number of studies.

The combined \textit{p}-value $p_E(\mu)$ serves as the basis for inference: It
can be used to quantify the evidence against any null hypothesis of interest,
and can be inverted to obtain point estimates and CIs. Specifically, a point
estimate is obtained by solving the equation $p_E(\mu) = 1/2$ for $\mu$, and a
$(1 - \alpha)\times 100\%$ CI by solving $p_E(\mu) = \alpha/2$ and $p_E(\mu) = 1
- \alpha/2$. For two studies, the point estimate has the closed form $\hat{\mu}
= (\hat{\theta}_1/\sigma_1 + \hat{\theta}_2/\sigma_2)/(1/\sigma_1 +
1/\sigma_2)$, that is, a weighted average of the effect estimates using
inverse-standard-error weights \citep{Pawel2025}. CI bounds and the point
estimate for more than two studies require numerical root-finding, which is
implemented in the \texttt{confMeta} R package \citep{Hofmann2026}.

Frequently, it is informative to display the \textit{p}-value function for a
range of null values $\mu$ in a so-called ``drapery plot'' \citep{Ruecker2020}.
For this purpose, it is natural to convert the combined \textit{p}-value
function based on one-sided \textit{p}-values into a two-sided combined
\textit{p}-value function via $\tilde{p}_E(\mu) = 2 \min\{p_E(\mu), 1 -
p_E(\mu)\}$ \citep{Xie2013, Held2025}. This function then peaks at the point
estimate and the bounds of a $(1 - \alpha)\times 100\%$ CI can be obtained by
cutting it at $\alpha$.

\begin{figure}[!htb]
\begin{knitrout}
\definecolor{shadecolor}{rgb}{0.969, 0.969, 0.969}\color{fgcolor}

{\centering \includegraphics[width=\maxwidth]{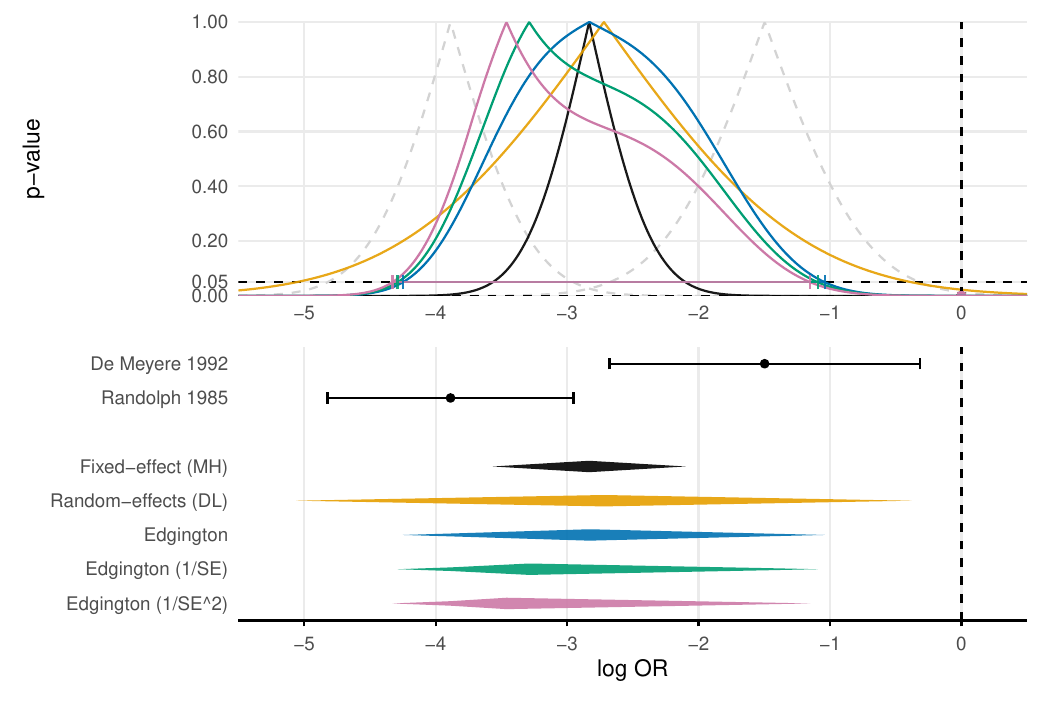} 

}

\end{knitrout}

\caption{Drapery plot (top) and forest plot (bottom) for meta-analysis of the
  effect of antibiotics compared to a placebo on sore throat
  \citep[Figure~2]{IQWiG2022}.}
\label{fig:draperyexample}
\end{figure}

Figure~\ref{fig:draperyexample} shows a drapery plot for the meta-analysis
example from the introduction (see again Figure~\ref{fig:metaexample} for
details). The \textit{p}-value function from Edgington's method (blue) has its
peak (and consequently its point estimate) at $\log \widehat{\text{OR}} =
-2.83$, which is identical to the fixed-effect
estimate $\log \widehat{\text{OR}} = -2.83$ based on the
Mantel-Haenszel (MH) method (black). Both point estimates lie between the
inverse-variance fixed-effect estimate $\log \widehat{\text{OR}} =
-2.96$ (not shown) and the DL random-effects estimate
$\log \widehat{\text{OR}} = -2.72$ (orange) based on the
PM estimate $\hat{\tau} = 1.6$ (95\% CI
from $0.52$ to
$10$). Despite sharing the same point
estimate, Edgington's method more explicitly reflects the observed between-study
discrepancy, with its 95\% CI ranging from $-4.25$ to $-1.04$, which is substantially wider
than the fixed-effect CI based on the MH method ($-3.57$ to
$-2.10$), but also considerably
narrower than the DL random-effects CI ($-5.06$ to
$-0.38$). Figure~\ref{fig:draperyexample} also shows
weighted versions of Edgington's method, which we introduce in the following
Section.

\subsection{Weighted Edgington's method}

The standard Edgington's method assigns equal weight to both studies. Since
studies typically differ in precision, it is natural to ask whether
incorporating study-specific weights can improve performance. Weighted
extensions of Edgington's method have previously been considered in the
replication study setting to ``downweight'' less trustworthy original studies
\citep{Held_etal2024}. In the meta-analysis context, this extension can also be
motivated by the fact that fixed-effect meta-analysis can be viewed as
\textit{p}-value combination via Stouffer's method with inverse-standard-error
weights \citep[p.6]{Pawel2025}. Stouffer's method is based on a weighted sum of
\textit{z}-statistics, whereas weighted extensions of Edgington's method are
based on a weighted sum of \textit{p}-values.

We define the weighted sum of \textit{p}-values as $s_w(\mu) = w_1 p_1(\mu) +
w_2 p_2(\mu)$. Assuming without loss of generality that $0 < w_1 \leq w_2$,
the combined \textit{p}-value is given by \citep[p.4]{Held_etal2024}
\begin{align*}
  p_{W}(\mu) =
  \begin{cases}
    \{s_w(\mu)\}^2 / (2 w_1 w_2) & \text{if} ~ 0 < s_w(\mu) \leq w_1 \\
    \left\{s_w(\mu) - w_1 / 2 \right\} / w_2 & \text{if} ~ w_1 < s_w(\mu) \leq w_2 \\
    1 - \left\{(w_1 + w_2) - s_w(\mu)\right\}^2 / (2 w_1 w_2)  & \text{if} ~ w_2 < s_w(\mu) \leq w_1 + w_2. \\
  \end{cases}
\end{align*}
Appendix~\ref{app:morethan2studies} shows the general form of the combined
\textit{p}-value from the weighted Edgington's method for meta-analysis of any
number of studies.

As in the unweighted case, point estimates and CI bounds are obtained by
numerically solving $p_W(\mu) = 1/2$, $p_W(\mu) = \alpha/2$, and $p_W(\mu) = 1 -
\alpha/2$ for $\mu$, respectively. There is no longer a closed-form solution for
the point estimate. In the following, we consider two weighting schemes:
inverse-standard-error ($w_i = 1/\sigma_i$) and inverse-variance ($w_i =
1/\sigma^2_i$) weights. As illustrated in Figure~\ref{fig:draperyexample},
introducing weights shifts the point estimate toward the more precise study,
while the weighted meta-analytic CIs remain similar to the CI from the
unweighted method.

\subsection{Comparing Edgington's method to fixed-effect meta-analysis}
\label{sec:Edgingtonconditions}
Both Edgington's method and fixed-effect meta-analysis are, by construction,
calibrated under homogeneity (i.e., their CIs have nominal coverage and combined
\textit{p}-values are uniformly distributed under the null hypothesis). For a
given set of standard errors $\sigma_1$ and $\sigma_2$, fixed-effect
meta-analysis using the inverse-variance method always gives a CI of the same
width irrespective of the point estimates, as the standard error of its
meta-analytic estimate, $\sqrt{1/(1/\sigma_1^2 + 1/\sigma_2^2)}$, depends only
on $\sigma_1$ and $\sigma_2$. In contrast, Edgington's method can produce a
wider CI than fixed-effect meta-analysis if there is discrepancy between the
study-specific point estimates. Since both methods have the same coverage under
homogeneity, there must also be situations in which Edgington's method produces
a narrower CI. We will now give sufficient conditions under which this is the
case.

Denote by $z_{q}$ the $q \times 100\%$ quantile of the standard normal
distribution, by $r = \sigma_1/\sigma_2$ the ratio of the first to the second
study's standard error, and by $Q = (\hat{\theta}_1 -
\hat{\theta}_2)^2/(\sigma^2_1 + \sigma^2_2)$ the $Q$-statistic for two studies.
Assuming $\alpha < 0.25$, the $(1 - \alpha) \times 100\%$ CI based on the
unweighted Edgington's method is narrower than the corresponding CI based on
inverse-variance fixed-effect meta-analysis if the following conditions hold
\begin{enumerate}
   \item[] $Q <
  \left\{z_{\sqrt{\alpha/4}}\,\frac{r + 1}{\sqrt{r^2 + 1}} - z_{\alpha/2} \, \frac{2
    \, r}{1 + r^2} \right\}^2$
  and
  $\max\left\{r^2, \frac{1}{r^2}\right\} <
  \frac{z_{\alpha/2}^2}{z_{\sqrt{\alpha/4}}^2} - 1$,
\end{enumerate}
see Appendix~\ref{app:proofwidth} for details and more general conditions for
the weighted Edgington's method. Formally, these conditions also require the
assumption $\sigma_1 \neq \sigma_2$, but the special case $\sigma_1 = \sigma_2$
is explained in Appendix~\ref{sec:bounds-edg-width}. These are only sufficient
but not necessary conditions. They nevertheless provide an intuitive
understanding of how Edgington's CI width depends on study discrepancy in terms
of $Q$ and the ratio of the study-specific standard errors $r$.

For example, if both studies have the same standard error ($r = 1$), Edgington's
95\% CI is guaranteed to be narrower than the fixed-effect CI if $Q <
0.06$, which corresponds to a $Q$-test \textit{p}-value
greater than $0.81$, indicating hardly any evidence against
homogeneity. If instead the first study's standard error is 20\% larger than the
second ($r = 1.2$), the condition is only satisfied for $Q <
0.05$ (equivalently, a $Q$-test \textit{p}-value greater
than $0.83$), so greater similarity between studies is
required when their standard errors are more imbalanced.

Similar sufficient conditions can be derived that guarantee that Edgington's CI
is wider than the fixed-effect CI, and these also depend only on $Q$, $r$, and
$\alpha$ (under the assumptions $\alpha < w_1/(4w_2)$ and $\sigma_1 \neq
\sigma_2$; see Appendix~\ref{app:proofwidth} for details).
These conditions yield explicit bounds on both $r$ and $Q$: If the $Q$-statistic
is greater than 0.78 (corresponding to a $Q$-test
\textit{p}-value less than $0.38$), the unweighted
Edgington's 95\% CI will always be wider than the fixed-effect CI regardless of
$r$. Similarly, if the standard error ratio satisfies $\max\{r, 1/r\} >
2.38$, the unweighted Edgington's CI is wider for any value
of $Q$.

In summary, to guarantee that Edgington's CI is narrower than the fixed-effect
CI, the observed between-study discrepancy in terms of $Q$ must be small and the
study-specific standard errors must be similar. Conversely, wider CIs are
guaranteed when $Q$ is large or the standard errors are sufficiently imbalanced.

\section{Application to IQWiG two-study meta-analyses}
\label{sec:iqwig}

We now apply the unweighted and weighted versions of Edgington's method to the
large database of two-study meta-analyses from IQWiG. The database lets us study
how the methods behave on real data and how they compare to established methods.

\subsection{Data descriptives}

The IQWiG database \citep{IQWiG2026} consists of 2290 meta-analyses of which 1226 contain
two studies only. Table~\ref{tab:IQWiG-descriptives} summarizes the effect
measures used across these two-study meta-analyses. Log risk ratios (RRs) and
log odds ratios (ORs) are the most common, followed by mean differences (MDs),
standardized mean differences (SMDs), and log hazard ratios (HRs).

\begin{table}[!h]
\centering
\caption{\label{tab:IQWiG-descriptives}Number of two-study meta-analyses in IQWiG database per effect measure along with the proportion of statistically significant fixed-effect (FE) meta-analytic effect estimates and $Q$-tests (both at $\alpha =
    0.05$).}
\centering
\fontsize{11}{13}\selectfont
\begin{tabular}[t]{lrrr}
\toprule
\textbf{Effect measure} & \textbf{Meta-analyses} & \textbf{FE sig. (\%)} & \textbf{$\boldsymbol{Q}$-test sig. (\%)}\\
\midrule
\cellcolor{Gray!10}{Log risk ratio (RR)} & \cellcolor{Gray!10}{316} & \cellcolor{Gray!10}{32.3} & \cellcolor{Gray!10}{4.4}\\
Log odds ratio (OR) & 304 & 29.3 & 5.3\\
\cellcolor{Gray!10}{Mean difference (MD)} & \cellcolor{Gray!10}{280} & \cellcolor{Gray!10}{39.3} & \cellcolor{Gray!10}{13.2}\\
Standardized mean difference (SMD) & 187 & 39.6 & 15.0\\
\cellcolor{Gray!10}{Log hazard ratio (HR)} & \cellcolor{Gray!10}{124} & \cellcolor{Gray!10}{34.7} & \cellcolor{Gray!10}{4.8}\\
Other & 15 & 26.7 & 20.0\\
\bottomrule
\end{tabular}
\end{table}

\begin{figure}[!htb]
\begin{knitrout}
\definecolor{shadecolor}{rgb}{0.969, 0.969, 0.969}\color{fgcolor}

{\centering \includegraphics[width=\maxwidth]{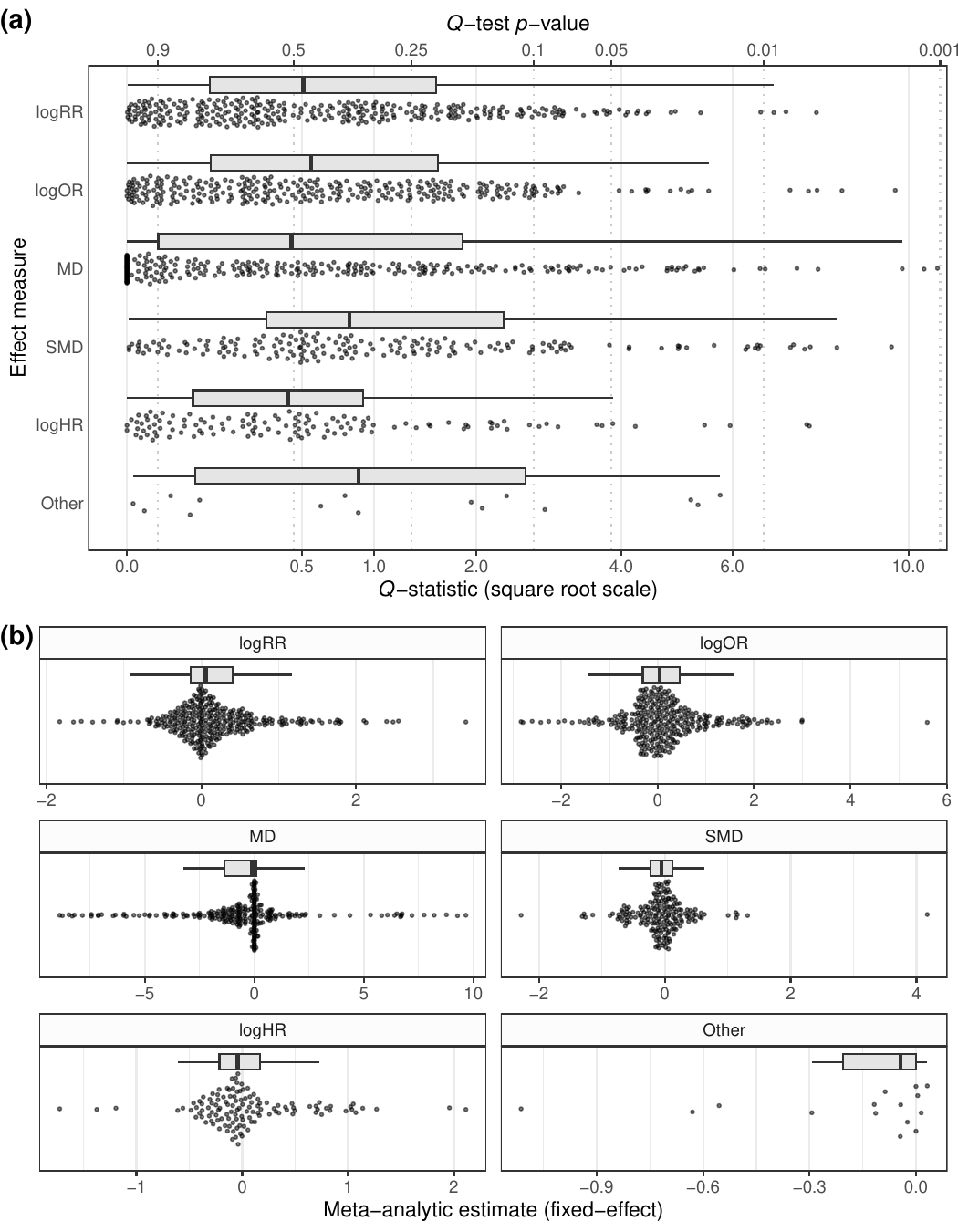} 

}

\end{knitrout}
\caption{Distribution of Cochran's $Q$ statistics (a) and fixed-effect
  estimates (b) across 1226
  IQWiG two-study meta-analyses. In plot (a), 24
  meta-analyses with $Q > 10$ are not shown. In plot (b),
  23 meta-analyses with MD estimates
  larger than 10 in absolute value are not shown.}
\label{fig:IQWiG-descriptives}
\end{figure}

We now examine fixed-effect meta-analysis results. Note that in actual IQWiG
reports, pooled meta-analysis results are usually not presented when
heterogeneity is deemed too large, while forest plots with study-level summaries
are always shown. The meta-analytic results below therefore include pooled
estimates that would not be presented in practice due to excessive
heterogeneity. Excluding the small ``Other'' category (e.g., ratio of means,
risk differences), the proportion of meta-analyses with a statistically
significant fixed-effect result (at $\alpha = 0.05$) roughly ranges from 30\% to
40\% across effect measure types, while the proportion with a significant
$Q$-test for homogeneity ranges from approximately 4\% to 15\%.

Figure~\ref{fig:IQWiG-descriptives} shows the distribution of $Q$-statistics (a)
and fixed-effect estimates (b) across all 1226
two-study meta-analyses. For two-study meta-analyses, the $Q$-statistic has a
$\chi^2_1$ distribution under homogeneity. The $Q$-statistic distributions are
relatively similar across effect measures, with most meta-analyses showing
little to moderate evidence against homogeneity, though the $Q$-test typically
has very low power with only two studies \citep{Hedges2001}. Across most effect
measures, fixed-effect estimates are mostly small in magnitude and roughly
symmetric around zero. MDs are an exception as this effect measure is not
standardized and therefore exhibits a wider range. The concentration of MD
estimates near zero likely reflects a substantial number of studies estimating
MDs on a small scale. Furthermore, the distribution of MD estimates (and to a
lesser extent logRRs and logORs) appears non-symmetric, with more meta-analyses
showing negative estimates. The reason could be that MDs in the IQWiG database
may be more often oriented in the way that negative MDs indicate treatment
benefit, and beneficial results may be more likely reported.

\subsection{Confidence interval comparison}

Table~\ref{tab:edgiqwig} summarizes characteristics of the 95\% CIs from
Edgington's methods alongside DL random-effects and HK random-effects
\citep[with ad hoc variance correction according to IQWiG's
  methodology;][]{IQWiG2025} CIs. Both random-effects methods use the PM
heterogeneity estimator, though the same results would be obtained with the REML
or DL estimators, as they all coincide for two-study meta-analyses
\citep{Jackson2017}.

In addition to correct coverage rate, it may be desirable for a meta-analytic CI
to possess other good properties. For instance, it may be desirable for a
meta-analytic CI to include both study-specific point estimates
\citep{Jackson2017}. Note that under a random-effects model, this property is
more naturally associated with a prediction interval, while under the
fixed-effect model the prediction interval coincides with the confidence
interval.
The CIs from Edgington's method (unweighted) include both study-specific point
estimates in all 1,226 meta-analyses (first
column), whereas the weighted variants do so in more than 90\% of meta-analyses
which is comparable to the random-effects methods. In contrast, fixed-effect
meta-analysis includes both point estimates in only about 75\% of cases, as its
CI cannot widen in response to between-study discrepancy.

\begin{table}[!h]
\centering
\caption{\label{tab:edgiqwig}Characteristics of meta-analytic 95\% CIs in IQWiG two-study meta-analyses. The first column gives the proportion of meta-analytic CIs that include both study-specific point estimates. The second column gives the proportion of meta-analytic CIs that are informative, i.e., do not extend beyond the range of the  study-specific CIs. The last two columns give the proportion of meta-analytic CIs that are narrower than both or only one study-specific CI, respectively.}
\centering
\fontsize{8.25}{10.25}\selectfont
\begin{tabular}[t]{lrrrr}
\toprule
\textbf{Method} & \textbf{Estimates included (\%)} & \textbf{Informative (\%)} & \textbf{Narrower both CIs (\%)} & \textbf{Narrower only one CI (\%)}\\
\midrule
\cellcolor{Gray!10}{Edgington} & \cellcolor{Gray!10}{100.0} & \cellcolor{Gray!10}{100.0} & \cellcolor{Gray!10}{62.6} & \cellcolor{Gray!10}{29.9}\\
Edgington (1/SE) & 98.1 & 100.0 & 69.7 & 23.4\\
\cellcolor{Gray!10}{Edgington (1/$\text{SE}^2$)} & \cellcolor{Gray!10}{92.1} & \cellcolor{Gray!10}{100.0} & \cellcolor{Gray!10}{76.3} & \cellcolor{Gray!10}{17.1}\\
Fixed-effect (IV) & 75.7 & 100.0 & 100.0 & 0.0\\
\cellcolor{Gray!10}{Fixed-effect (IV + MH)} & \cellcolor{Gray!10}{75.4} & \cellcolor{Gray!10}{100.0} & \cellcolor{Gray!10}{99.8} & \cellcolor{Gray!10}{0.2}\\
Random-effects (DL) & 96.0 & 90.6 & 72.1 & 12.4\\
\cellcolor{Gray!10}{Random-effects (HK)} & \cellcolor{Gray!10}{99.7} & \cellcolor{Gray!10}{6.8} & \cellcolor{Gray!10}{2.1} & \cellcolor{Gray!10}{6.4}\\
\bottomrule
\end{tabular}
\end{table}

It may also be desirable for a meta-analytic CI to be ``informative'' in the
sense that the CI does not extend beyond the range spanned by the study-specific
CIs \citep{Schulz2021}. Looking at the proportion of informative CIs (second
column in Table~\ref{tab:edgiqwig}), all Edgington variants and fixed-effect
meta-analysis produce informative CIs in every meta-analysis. This is not the
case for DL random-effects meta-analysis which shows an informative CI in
$90.6\%$ of meta-analyses. HK random-effects meta-analysis leads to
an informative CI in only $6.8\%$ of meta-analyses, illustrating
clearly its limited practical value in the two-study setting.

With respect to CI width, inverse-variance fixed-effect meta-analysis always
yields meta-analytic CIs narrower than both study-specific CIs (last two columns
in Table~\ref{tab:edgiqwig}), as expected from the theoretical properties of the
method. With the MH fixed-effect method for $2 \times 2$ tables, which IQWiG
uses, a small proportion of meta-analyses have a meta-analytic CI narrower than
only one study-specific CI. The Edgington and DL random-effects methods produce
meta-analytic CIs narrower than both study-specific CIs in roughly 60\% to 75\%
of the cases, whereas HK random-effects meta-analysis achieves narrower CIs in
only $2.1\%$ of the meta-analyses.

\begin{figure}[!htb]
\begin{knitrout}
\definecolor{shadecolor}{rgb}{0.969, 0.969, 0.969}\color{fgcolor}

{\centering \includegraphics[width=\maxwidth]{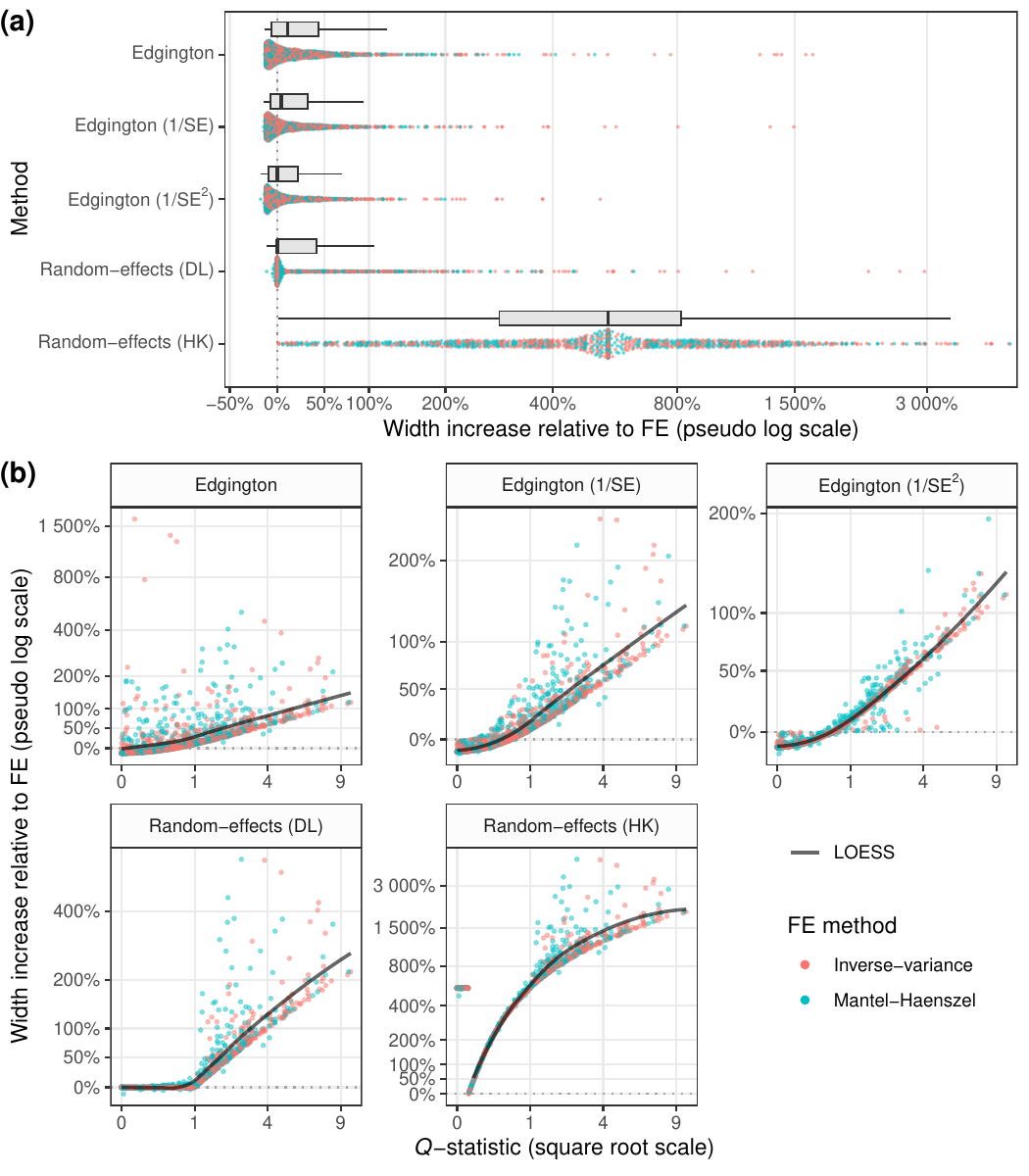} 

}

\end{knitrout}
\caption{CI comparisons of different meta-analytic methods with fixed-effect
  (FE) meta-analysis. Plot (a) shows the distribution of relative 95\% CI width
  increase, $(\text{width} - \text{width}_{\text{FE}})/\text{width}_{\text{FE}}
  = \text{width}/\text{width}_{\text{FE}} - 1$, for different methods (for HK,
  excluding 14 meta-analyses where the relative increase is
  greater than 4000\%). Plot (b) shows relative
  increase as a function of the $Q$-statistic with superimposed LOESS smoother
  (fitted on both IV and MH FE meta-analysis, but excluding
  24 extreme meta-analyses with $Q >
  10$, and, for HK, excluding meta-analyses with $Q <
  0.05$ to avoid fitting the bump caused by the ad hoc
  correction). Relative width increase is shown on a pseudo-log scale
  \citep{Wickham2025b} to compress large values while preserving detail around
  zero.}
\label{fig:iqwigmethodcomparison}
\end{figure}

Figure~\ref{fig:iqwigmethodcomparison}a shows the distribution of the CI width
increase relative to fixed-effect meta-analysis. HK random-effects meta-analysis
produces the widest CIs by far. In some meta-analyses, the relative CI width
increase is more than 3000\%. The median relative width increase of
$548\%$ reflects the ad hoc
HK correction used by IQWiG, which switches to the DL random-effects standard
error whenever the HK method would otherwise yield a narrower CI than
DL\footnote{This value of $548\%$ (plus $100\%$ since relative width increase is here defined as
$\text{width}/\text{width}_{\text{FE}} - 1$) corresponds to the ratio of the
97.5\% standard Cauchy to the 97.5\% standard normal quantile, the critical
values used to construct 95\% CIs under HK and FE respectively. The
concentration of the distribution at exactly this characteristic ratio follows
from three factors: (i) IQWiG's ad hoc correction switches to the DL
random-effects standard error (with PM estimator) whenever HK (with PM
estimator) would yield a narrower CI than DL random-effects (with DL
heterogeneity estimator); (ii) only two-study meta-analyses are considered here,
where the PM and DL heterogeneity estimators coincide and admit a closed-form
expression $\hat{\tau}^2 = \max\{Q - 1, 0\}(\sigma^2_1 + \sigma^2_2)/2$
\citep[Section~5.5]{Jackson2017}; (iii) for all such meta-analyses,
$\hat{\tau}^2 = 0$ as $Q \leq 1$, so the DL random-effects standard error
coincides with the inverse-variance FE standard error, hence the only difference
between HK and FE being the critical value used to construct the CI.}
\citep{IQWiG2025}. The distributions of the relative increase from the other
methods are all right-skewed and closer to 0\% than that of HK. All Edgington
variants can produce narrower CIs than fixed-effect meta-analysis, i.e., can
show a negative relative increase. This is not possible under DL random-effects
meta-analysis, which can only widen or maintain the fixed-effect CI width. The
only exception is that MH fixed-effect CIs (blue points) can be wider than DL
random-effects CIs. The most extreme widening among Edgington's methods occurs
for the unweighted variant, as this method here always includes both
study-specific point estimates even in cases of extreme discrepancy. The
inverse-standard-error weighted variant shows more moderate behavior, while the
inverse-variance weighted variant has a median relative increase of zero with
substantially less extreme widening overall.

Figure~\ref{fig:iqwigmethodcomparison}b shows the relative CI width increase as
a function of the $Q$-statistic. For small values of $Q$ (i.e., little observed
discrepancy), Edgington's methods are more likely to produce narrower CIs than
fixed-effect meta-analysis. As $Q$ increases, the relative CI width of all
Edgington variants also increases, with the inverse-variance weighted variant
increasing the least followed by the inverse-standard-error and the unweighted
variant. In contrast, the DL random-effects CI does not substantially widen
relative to the fixed-effect CI until around $Q > 1$, but then widens more
steeply (relative to the fixed-effect CI) than any of the Edgington variants.
This is expected as the DL/PM/REML estimator $\hat{\tau}^2 = \max\{Q - 1,
0\}(\sigma^2_1 + \sigma^2_2)/2$ \citep[Section~5.5]{Jackson2017} produces a
positive heterogeneity estimate only when $Q > 1$. Overall, Edgington's methods
occupy a middle ground between fixed-effect and DL random-effects meta-analysis:
they adapt to discrepancy more gradually than DL, and can even produce narrower
CIs than fixed-effect meta-analysis when discrepancy is low.
Figure~\ref{fig:iqwigCIcomparisonApp} in Appendix~\ref{app:CIwidthresults} shows
this at an aggregated level via the proportion of meta-analyses with a relative
CI width increase greater than or equal to 0\% stratified by $Q$-statistic.

\subsection{Point estimate comparison}

Edgington's method produces not only a different CI but also a different point
estimate than standard meta-analysis methods. We therefore now compare point
estimates from Edgington's and the random-effects methods with the fixed-effect
meta-analysis point estimate.

\begin{figure}[!htb]
\begin{knitrout}
\definecolor{shadecolor}{rgb}{0.969, 0.969, 0.969}\color{fgcolor}

{\centering \includegraphics[width=\maxwidth]{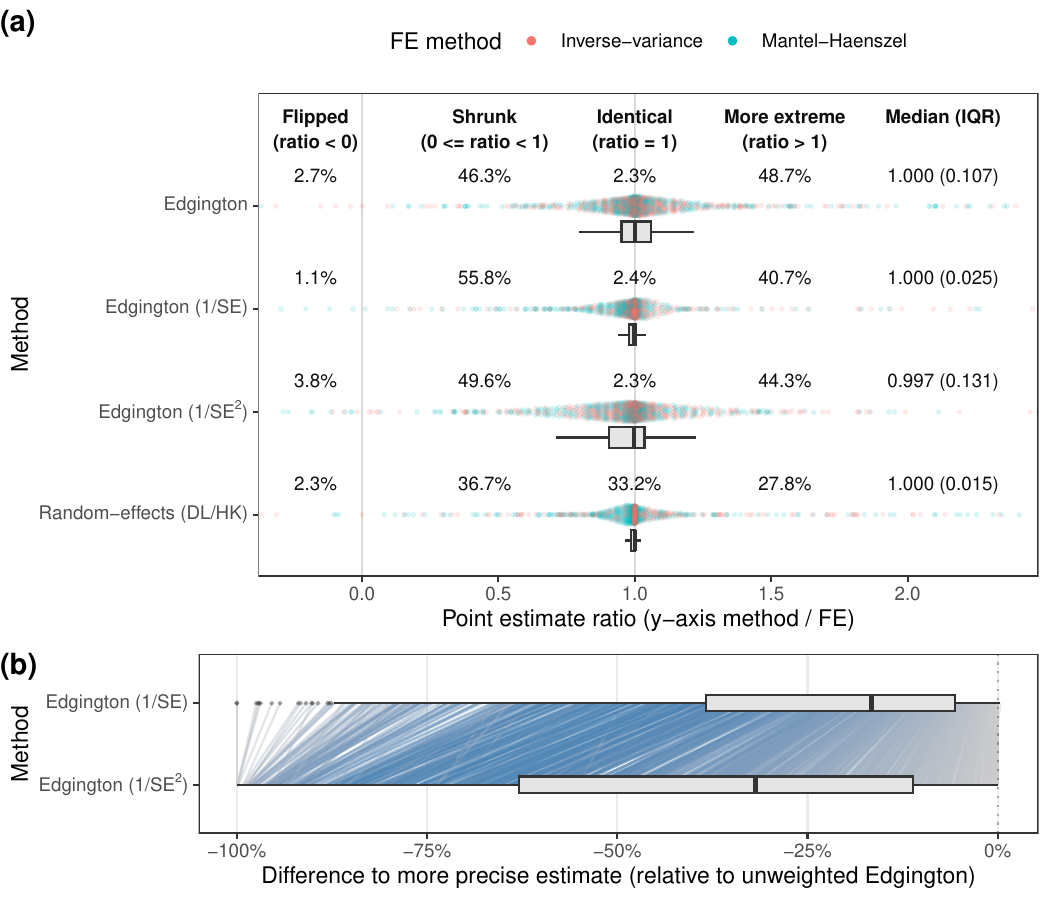} 

}

\end{knitrout}
\caption{Point estimate comparisons of different meta-analytic methods. Plot~(a)
  shows the distribution of the ratio of the point estimate of each method to
  that of fixed-effect meta-analysis (FE), along with the percentage of
  meta-analyses in which the method's estimate showed a flipped sign, was
  shrunk, was identical, or more extreme in the same direction compared to FE.
  Ratios outside the plot range are not shown but are included in the
  percentages. 41 meta-analyses where the fixed-effect estimate
  is within a tolerance of $10^{-6}$ around zero are
  excluded to avoid division by zero, and the same tolerance around a ratio of
  one is used to classify identical point estimates. Plot~(b) shows, for the
  weighted Edgington variants, the absolute difference between the meta-analytic
  point estimate and the more precise study-specific estimate relative to the
  corresponding difference for the unweighted variant
  ($(\text{diff}_{\text{weighted}} - \text{diff}_{\text{unweighted}}) /
  \text{diff}_{\text{unweighted}}$). Negative values indicate that the weighted
  variant is closer to the more precise study-specific estimate than the
  unweighted variant. The darkness of the color of the connecting lines reflects
  the difference between the two weighted variants. 51
  meta-analyses where the estimate from Edgington's method is equal to the more
  precise study-specific estimate are excluded to avoid division by zero.}
\label{fig:pointestimate}
\end{figure}

Figure~\ref{fig:pointestimate}a shows the distribution of the ratio of the point
estimate of each method to that of fixed-effect meta-analysis. The ratio is
useful because it standardizes different effect measure types to a common scale,
and makes it possible to assess whether the direction of the estimate changed
relative to fixed-effect meta-analysis (a flipped sign; ratio $<$ 0), the
estimate was shrunken toward zero (0 $\leq$ ratio $<$ 1), was identical (ratio
$=$ 1), or was more extreme in the same direction (ratio $>$ 1). For all
methods, the median ratio is close to one, meaning that roughly half of the
meta-analyses show a smaller and half a larger point estimate than fixed-effect
meta-analysis. Flipped signs are rare across all methods, occurring in fewer
than 4\% of meta-analyses. Both random-effects methods produce identical point
estimates, as HK and DL differ only in how the CI is constructed. Both also
coincide with the fixed-effect estimate in around a third of meta-analyses,
reflecting that random-effects methods reduce to fixed-effect meta-analysis
whenever heterogeneity is estimated as zero. The Edgington variants show fewer
identical estimates since they do not reduce to fixed-effect meta-analysis in
the absence of heterogeneity. The inverse-standard-error weighted variant shows
the fewest flipped signs among all methods (1.1\%) and produces point estimates that are the closest and least dispersed
relative to fixed-effect meta-analysis among the three Edgington variants
(interquartile range of 0.025). The unweighted and
inverse-variance weighted variants instead show considerably wider dispersion
(interquartile ranges 0.107 and 0.131, respectively).

To better understand the effect of the weights on Edgington's method,
Figure~\ref{fig:pointestimate}b examines how the point estimates of the weighted
Edgington variants relate to those of the unweighted variant. In all
meta-analyses, the weighted Edgington variants produce point estimates closer to
the more precise study-specific estimate than the unweighted variant. The
inverse-variance weighted variant shows a larger shift than the
inverse-standard-error weighted variant, consistent with its stronger
downweighting of the less precise study. This may also explain the dispersion
pattern in Figure~\ref{fig:pointestimate}a: the unweighted variant ignores
precision entirely, while inverse-variance weighting downweights the less
precise study most strongly. Inverse-standard-error weighting falls in between,
keeping point estimates closer to fixed-effect meta-analysis.

\subsection{Statistical significance comparison}

\begin{figure}[!htb]
\begin{knitrout}
\definecolor{shadecolor}{rgb}{0.969, 0.969, 0.969}\color{fgcolor}

{\centering \includegraphics[width=\maxwidth]{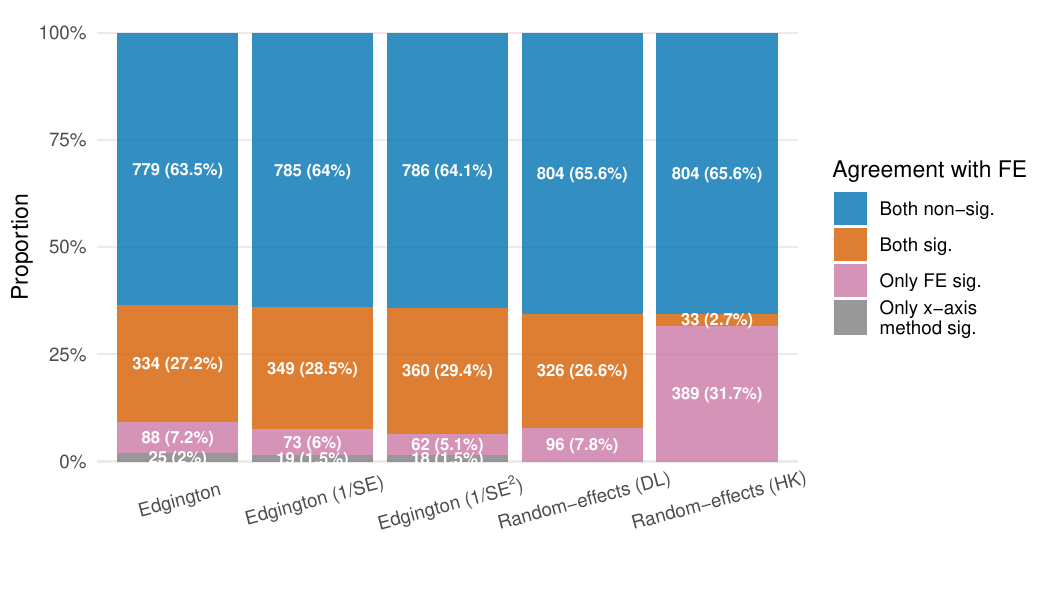} 

}

\end{knitrout}
\caption{Statistical significance agreement between different meta-analytic
  methods and fixed-effect (FE) meta-analysis at two-sided $\alpha = 0.05$.}
\label{fig:iqwigsigagree}
\end{figure}

Figure~\ref{fig:iqwigsigagree} shows the statistical significance agreement (at
$\alpha = 0.05$) between each method and fixed-effect meta-analysis. In the
large majority of meta-analyses, fixed-effect and all Edgington's methods reach
the same conclusions: both significant or both non-significant. Disagreements
occur in only $9.2\%$ of meta-analyses for unweighted Edgington,
decreasing slightly for the weighted variants. Disagreements arise in both
directions: in some cases fixed-effect is significant while Edgington is not,
and in a smaller number of cases, an Edgington's method is significant while
fixed-effect is not. By contrast, the random-effects methods never reach
statistical significance when fixed-effect does not, with HK yielding hardly any
significant results at all.

\begin{figure}[p]
\begin{knitrout}
\definecolor{shadecolor}{rgb}{0.969, 0.969, 0.969}\color{fgcolor}

{\centering \includegraphics[width=\maxwidth]{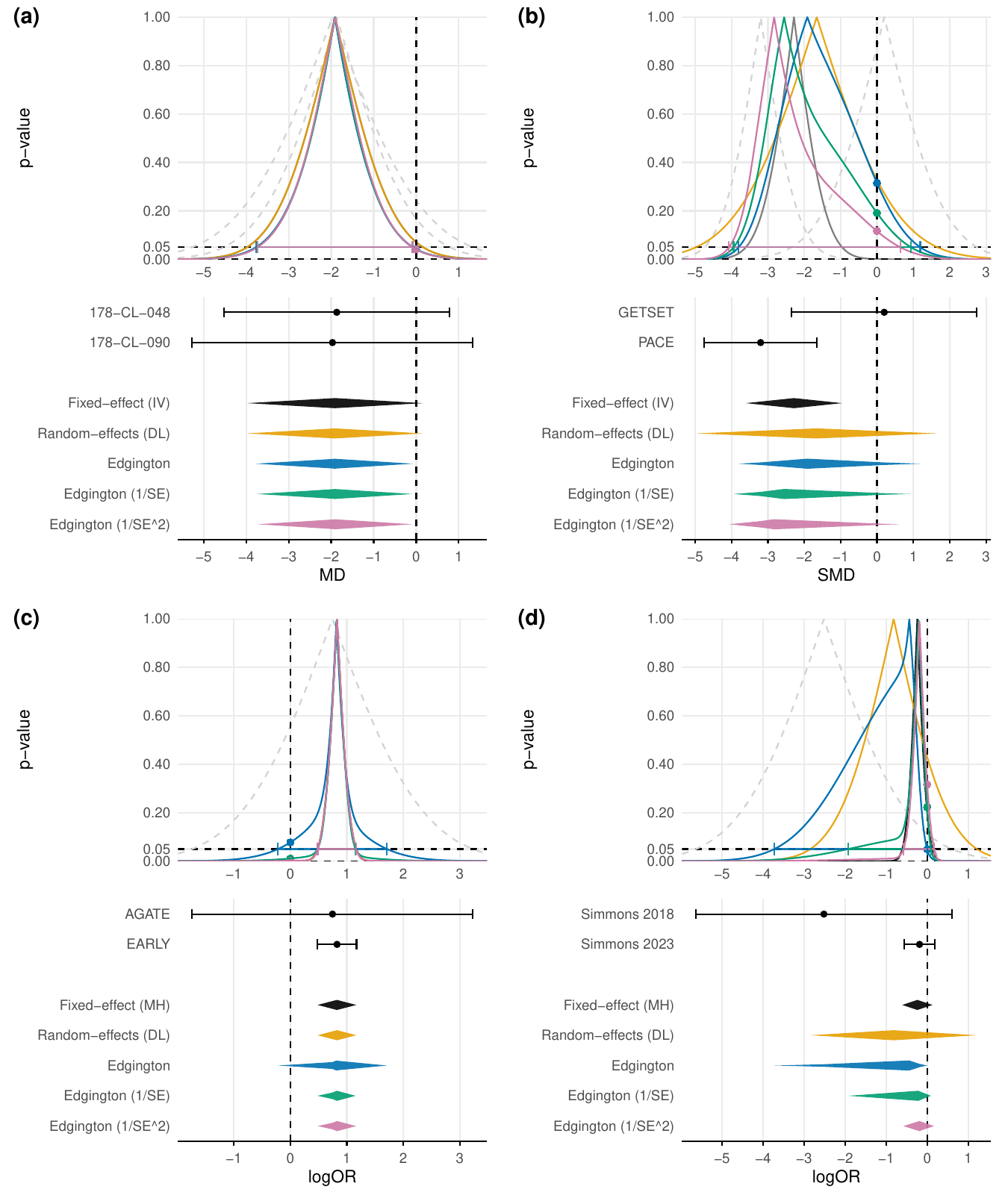} 

}

\end{knitrout}
\caption{Four examples of IQWiG two-study meta-analyses where fixed-effect
  meta-analysis and at least one of the Edgington's methods lead to different
  statistical significance conclusions. Plot (a) illustrates that CIs based on
  Edgington's method can be narrower for homogeneous study results. Plot (b)
  illustrates that Edgington's methods react to discrepancy in study results and
  the corresponding CIs can be asymmetric. The bottom plots illustrate how
  differing study-specific standard errors affect the CIs based on Edgington's
  methods for studies with similar~(c) and differing point estimates~(d).}
\label{fig:discrepantexamples}
\end{figure}

Table~\ref{tab:conflict} in Appendix~\ref{app:significance} lists all
meta-analyses with significance disagreement.
Figure~\ref{fig:discrepantexamples} illustrates four representative examples of
significance discrepancy.
Figure~\ref{fig:discrepantexamples}a shows that if
study results are highly homogeneous, Edgington's CI can be narrower than the
fixed-effect CI and shift the conclusion to significance. In
Figure~\ref{fig:discrepantexamples}b, the study results are strongly discrepant
so that Edgington's CIs widen and the conclusion becomes non-significant. In
Figure~\ref{fig:discrepantexamples}c, the CI of the unweighted method remains
very wide because of the substantially different standard errors of the studies,
despite the study point estimates being virtually identical. The CIs from the
weighted variants, on the other hand, largely coincide with fixed-effect and DL
random-effects CIs in this case. Finally, in
Figure~\ref{fig:discrepantexamples}d, strong discrepancy in both study-specific
point estimates and standard errors leads the unweighted Edgington's method to
produce an extremely wide CI, due to the method's tendency to include both point
estimates. The inverse-variance weighting variant concentrates its CI on the
more precise study, while the inverse-standard-error variant provides a
compromise between the two.

\section{Discussion}
This paper investigated Edgington's combination method for two-study
meta-analysis, motivated by the well-known drawbacks of random-effects methods
in this setting. Applying the method to 1226 meta-analyses from IQWiG, we found
that it exhibits a practically useful adaptive behavior: CIs widen in response
to observed between-study discrepancy and narrow when study results are highly
consistent, without requiring explicit estimation of a heterogeneity parameter
and with calibration under homogeneity. This places the method between standard
fixed-effect and random-effects approaches, with inferences coinciding with
fixed-effect meta-analysis in the large majority of cases while remaining
responsive to observed heterogeneity.

The unweighted Edgington's method can produce very wide intervals in cases of
extreme discrepancy (both with respect to study-specific point estimates and
standard errors). Weighted extensions of Edgington's method provide a natural
way of incorporating study precision, shifting point estimates toward the more
precise study while preserving the method's adaptive CI behavior to some extent.
The inverse-standard-error weighting offers a useful compromise between the
unweighted version and the more aggressively weighted inverse-variance scheme,
and may be preferable when studies differ substantially in precision.

Several limitations deserve mention. Our empirical investigation is based on a
single database, and results may differ in other evidence synthesis contexts,
such as Cochrane reviews or pharmaceutical regulatory submissions, where study
designs, effect measure types, and other characteristics may vary. Furthermore,
the current work focuses on descriptive comparisons of empirical data rather
than a formal simulation study under controlled conditions with known ground
truth; a dedicated simulation study assessing operating characteristics (e.g.,
CI coverage, bias, and type-I error rate) across realistic settings varying
degrees of effect size, heterogeneity, and study size imbalance would strengthen
the evidence base for adopting the method. Nevertheless, the simulation study by
Held et al. \citep{Held2025} has already demonstrated good properties of
Edgington's (unweighted) method in the case of three studies. Finally, the
choice of weights for the weighted Edgington's method is largely heuristic, and
principled guidance on weight selection requires further empirical and
theoretical investigations.

Taken together, our results indicate that Edgington's method is a practically
useful alternative to fixed-effect meta-analysis in the two-study setting,
occupying a middle ground between fixed-effect and random-effects approaches.
Before Edgington's method or one of its weighted extensions can replace the
standard fixed-effect method in future IQWiG meta-analyses, a systematic
simulation study is required, and such a study should also inform the choice of
weights.

\subsection*{Author contributions}
Conceptualization: SP, SF, LS, FW, GS, RB, LH; Data curation: SS, LS, SF, SP;
Formal analysis: SP, SF; Funding acquisition: LH, RB; Investigation: SP, SF, LH;
Methodology: SP, LH, SF, JC; Project administration: RB; Resources: LH;
Software: SF, SP; Supervision: SP, LH; Validation: SF, SP; Visualization: SF,
SP; Writing -- original draft: SP, SF, LH; Writing -- review \& editing: all

\subsection*{Acknowledgments}
We thank Fadoua Balabdaoui for support with derivations. We used Claude
\citep{Claude2026} for language and grammar checking and for reviewing and
debugging R code during manuscript preparation. We reviewed, revised, and
approved all AI-assisted content and take full responsibility for the final
manuscript.

\subsection*{Conflict of interest}
The authors declare no competing interests.

\subsection*{Funding}
This research received no specific funding.

\subsection*{Software and data}
The data from IQWiG were extracted from the forest plots in all published
reports and are available at
\url{https://www.iqwig.de/download/data_two_studies.xlsx}. Code to reproduce our
analyses is available at \url{https://gitlab.uzh.ch/crsuzh/edgington-iqwig}. A
snapshot of the repository at the time of writing is available at
\url{https://doi.org/10.5281/zenodo.21620885}. We used the statistical
programming language R version 4.6.1 (2026-06-24) for analyses
\citep{R} along with the \texttt{confMeta} \citep{Hofmann2026}, \texttt{metafor}
\citep{Viechtbauer2010}, \texttt{ggplot2} \citep{Wickham2016}, \texttt{dplyr}
\citep{Wickham2026}, \texttt{tidyr} \citep{Wickham2025}, \texttt{readxl}
\citep{Wickham2026b}, \texttt{ggbeeswarm} \citep{Clarke2025}, \texttt{ggpubr}
\citep{Kassambara2025}, \texttt{kableExtra} \citep{Zhu2024}, and \texttt{knitr}
\citep{Xie2015} packages.

{
\onehalfspacing 
\bibliographystyle{ama-doi}
\bibliography{bibliography}
}

\begin{appendices}
\renewcommand\thefigure{\thesection\arabic{figure}}
\renewcommand\thetable{\thesection\arabic{table}}

\section{The R package confMeta}
\label{app:package}
\setcounter{figure}{0}
\setcounter{table}{0}

The \texttt{confMeta} R package can be installed by running
\texttt{install.packages("confMeta")} in an R console. The following code
excerpt shows how \texttt{confMeta} can be used to conduct a meta-analysis with
Edgington's method using the example from Figure~\ref{fig:discrepantexamples}b.

\begin{spacing}{1}
\begin{knitrout}\small
\definecolor{shadecolor}{rgb}{0.969, 0.969, 0.969}\color{fgcolor}\begin{kframe}
\begin{alltt}
\hlkwd{library}\hldef{(confMeta)} \hlcom{# load package}

\hlcom{## inputs: effect estimates, standard errors, and study labels}
\hldef{yi} \hlkwb{<-} \hlkwd{c}\hldef{(}\hlnum{0.2}\hldef{,} \hlopt{-}\hlnum{3.2}\hldef{)}
\hldef{sei} \hlkwb{<-} \hlkwd{c}\hldef{(}\hlnum{1.3}\hldef{,} \hlnum{0.79}\hldef{)}
\hldef{labels} \hlkwb{<-} \hlkwd{c}\hldef{(}\hlsng{"GETSET"}\hldef{,} \hlsng{"PACE"}\hldef{)}

\hlcom{## conduct meta-analyses with Edgington's method}
\hlcom{## specify PM estimator for DL-RE comparison method in plots}
\hldef{cm_edgington} \hlkwb{<-} \hlkwd{confMeta}\hldef{(}\hlkwc{estimates} \hldef{= yi,} \hlkwc{SEs} \hldef{= sei,} \hlkwc{study_names} \hldef{= labels,}
                         \hlkwc{fun} \hldef{= p_edgington,} \hlkwc{fun_name} \hldef{=} \hlsng{"Edgington"}\hldef{,}
                         \hlkwc{method.tau.re} \hldef{=} \hlsng{"PM"}\hldef{)}
\hldef{cm_edgington}
\end{alltt}
\begin{verbatim}
##   Meta-Analysis with p-value combination
## -----------------------------------------
## Number of studies:  2 
## Combination method: Edgington 
## 
## Estimate: -1.915,
## 95% Confidence Interval: [-3.803, 1.188]
\end{verbatim}
\begin{alltt}
\hlcom{## with inverse-standard-error weighted Edgington's method}
\hldef{cm_edgington_w} \hlkwb{<-} \hlkwd{confMeta}\hldef{(}\hlkwc{estimates} \hldef{= yi,} \hlkwc{SEs} \hldef{= sei,} \hlkwc{study_names} \hldef{= labels,}
                           \hlkwc{fun} \hldef{= p_edgington_w,} \hlkwc{fun_name} \hldef{=} \hlsng{"Edgington (1/SE)"}\hldef{,}
                           \hlkwc{w} \hldef{=} \hlnum{1}\hlopt{/}\hldef{sei)}
\hldef{cm_edgington_w}
\end{alltt}
\begin{verbatim}
##   Meta-Analysis with p-value combination
## -----------------------------------------
## Number of studies:  2 
## Combination method: Edgington (1/SE) 
## 
## Estimate: -2.553,
## 95% Confidence Interval: [-3.942, 0.931]
\end{verbatim}
\begin{alltt}
\hlcom{## create drapery and forest plot with FE and RE (DL) meta-analysis comparison}
\hlkwd{autoplot}\hldef{(cm_edgington, cm_edgington_w,} \hlkwc{xlab} \hldef{=} \hlsng{"SMD"}\hldef{,}
         \hlkwc{reference_methods_p} \hldef{=} \hlkwd{c}\hldef{(}\hlsng{"re"}\hldef{,} \hlsng{"fe"}\hldef{),}
         \hlkwc{reference_methods_forest} \hldef{=} \hlkwd{c}\hldef{(}\hlsng{"re"}\hldef{,} \hlsng{"fe"}\hldef{),}
         \hlkwc{ref_labels} \hldef{=} \hlkwd{c}\hldef{(}\hlsng{"fe"} \hldef{=} \hlsng{"Fixed-effect (IV)"}\hldef{,}
                        \hlsng{"re"} \hldef{=} \hlsng{"Random-effects (DL-PM)"}\hldef{))}
\end{alltt}
\end{kframe}

{\centering \includegraphics[width=\maxwidth]{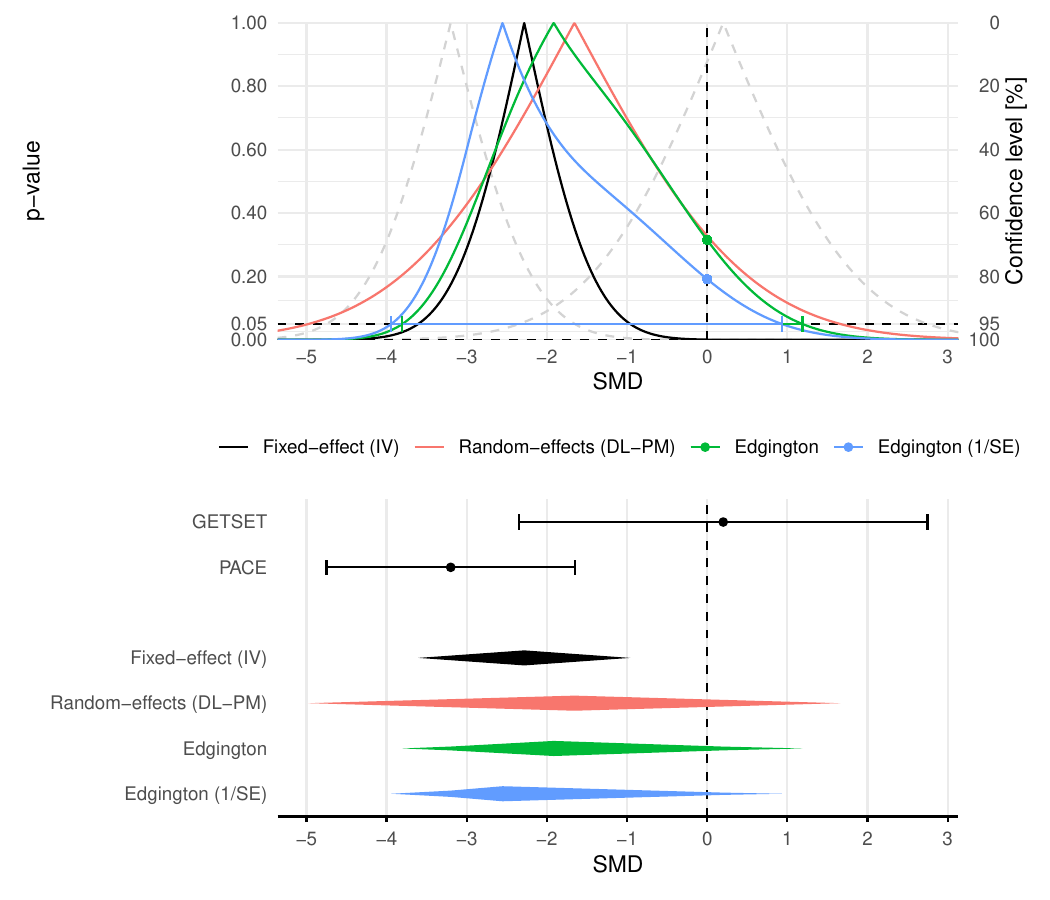} 

}

\end{knitrout}
\end{spacing}

\section{More than two studies}
\label{app:morethan2studies}
\setcounter{figure}{0}
\setcounter{table}{0}

For a meta-analysis of $k$ studies, Edgington's combination method is based on
the sum of the \textit{p}-values $s(\mu) = \sum_{i=1}^k p_i(\mu)$. A combined
\textit{p}-value can be computed from the cumulative distribution function of
the Irwin-Hall distribution with parameter $k$ by
\begin{align*}
  p_E(\mu) = \frac{1}{k!} \sum_{j=0}^{\lfloor s(\mu) \rfloor} (-1)^j \binom{k}{j}(s(\mu) - j)^k,
\end{align*}
where $\lfloor s(\mu) \rfloor$ denotes the greatest integer less than or equal
to $s(\mu)$. 
Similarly, the weighted Edgington's combination method is based on the sum of
weighted \textit{p}-values $s_w(\mu) = \sum_{i=1}^k w_i \, p_i(\mu)$. Using
results from Barrow and Smith \citep{Barrow1979} on the distribution of the sum
of weighted uniform random variables, the corresponding combined
\textit{p}-value is given by
\begin{align*}
  p_{W}(\mu) = \frac{1}{k! \prod_{i=1}^k w_i} \sum_{\boldsymbol{v} \in \{0,1\}^k}
  \mathbb{1}(s_w(\mu) - \boldsymbol{w}^\top \boldsymbol{v} \geq 0) (-1)^{\sum_{j=1}^k
    v_j} (s_w(\mu) - \boldsymbol{w}^\top \boldsymbol{v})^k
\end{align*}
where $\boldsymbol{w} = (w_1, \dots, w_k)^\top$ and $\mathbb{1}(\cdot)$ is the
indicator function. 

\section{Conditions for narrower/wider CIs}
\label{app:proofwidth}
\setcounter{figure}{0} \setcounter{table}{0}

This Appendix gives technical details on the derivation of conditions for
Edgington's method to produce narrower or wider CIs than fixed-effect
meta-analysis (using the inverse-variance method).

\subsection{Bounds on Edgington's CI width}
\label{sec:bounds-edg-width}

For weighted (and unweighted) Edgington's method, one needs to solve equations
of the form
\begin{align*}
    w_1 \Phi\left(\frac{\mu - \hat{\theta}_1}{\sigma_1}\right) + w_2 \Phi\left(\frac{\mu - \hat{\theta}_2}{\sigma_2}\right) = A
\end{align*}
for $\mu$ to obtain the lower and upper limit of a CI, where $A$ is a constant
depending on the confidence level $(1-\alpha)$ and whether the lower or upper
limit is of interest. In general, a closed-form solution does not exist. In the
following, we show a possible way to bound the lower and upper limits of
Edgington's CI under the assumption $\alpha < w_1/(4w_2)$ and $\sigma_1 \neq
\sigma_2$ (for the case $\sigma_1 = \sigma_2$, see further below).

\subsubsection{Well-separation}

\begin{sloppypar}
We now introduce the concept of well-separation to determine which of $p_1(\mu)
= \Phi\left\{(\mu - \hat{\theta}_1)/\sigma_1\right\}$ and $p_2(\mu) =
\Phi\left\{(\mu - \hat{\theta}_2)/\sigma_2\right\}$ is larger. This comparison
enables us to bound their sum and express it in terms of a single cumulative
distribution function.
\end{sloppypar}

Denote the point of intersection between the two individual $p$-value functions
$p_1(\mu)$ and $p_2(\mu)$ by $\mu^\ast$, which satisfies
\begin{align*}
  \Phi\left(\frac{\mu^\ast - \hat{\theta}_1}{\sigma_1}\right)
  = \Phi\left(\frac{\mu^\ast - \hat{\theta}_2}{\sigma_2}\right)
\end{align*}
and can be obtained by (assuming $\sigma_1 \neq \sigma_2$)
\begin{align*}
  \mu^\ast = \frac{\hat{\theta}_1/\sigma_1 - \hat{\theta}_2/\sigma_2}{1/\sigma_1 - 1/\sigma_2}.
\end{align*}
For $\sigma_1 = \sigma_2$, $p_1(\mu)$ and $p_2(\mu)$ never intersect (if
$\hat{\theta}_1 \neq \hat{\theta}_2$) or they coincide for all $\mu$ (if
$\hat{\theta}_1 = \hat{\theta}_2$). The weighted sum of the \textit{p}-value
functions evaluated at $\mu^\ast$ is then
\begin{align}
  \label{eq:sumatintersection}
  w_1 p_1(\mu^\ast) + w_2p_2(\mu^\ast) = \left(w_1+w_2\right) \Phi\left(\frac{\hat{\theta}_1 - \hat{\theta}_2}{\sigma_2 - \sigma_1}\right).
\end{align}

Assuming that\footnote{For the unweighted Edgington's method, this condition
holds for any $\alpha < 1$, whereas for the weighted variant, it holds for
$\alpha = 0.05$ and typical weight choices (e.g., inverse-standard-error or
inverse-variance) if the study-specific standard errors are not too different.}
$\alpha < w_1/w_2$, we call $p_1(\mu)$ and $p_2(\mu)$ \emph{well-separated} at
level $\alpha$, if $p_{W}(\mu^\ast) \leq \alpha/2$ or $p_{W}(\mu^\ast) \geq 1 -
\alpha/2$. Note that, for $\alpha < w_1/w_2$, a combined \textit{p}-value
$p_{W}(\mu^\ast) \leq \alpha/2 < w_1/(2w_2)$ is equivalent to $w_1 p_1(\mu^\ast)
+ w_2 p_2(\mu^\ast) \leq \sqrt{w_1w_2}\sqrt{\alpha}$, which is equivalent to
$(\hat{\theta}_1 - \hat{\theta}_2)/(\sigma_2 - \sigma_1) \leq
z_{\frac{\sqrt{w_1w_2}}{w_1+w_2}\sqrt{\alpha}}$. Similarly, a combined
\textit{p}-value $p_{W}(\mu^\ast) \geq 1-\alpha/2 > 1- w_1/(2w_2)$ is equivalent
to $w_1 p_1(\mu^\ast) + w_2 p_2(\mu^\ast) \geq
(w_1+w_2)-\sqrt{w_1w_2}\sqrt{\alpha}$, which is equivalent to $(\hat{\theta}_1 -
\hat{\theta}_2)/(\sigma_2 - \sigma_1) \geq
-z_{\frac{\sqrt{w_1w_2}}{w_1+w_2}\sqrt{\alpha}}$. Thus, $p_1(\mu)$ and
$p_2(\mu)$ are well-separated at level $\alpha$ if and only if
\begin{align*}
  \bigg\lvert\frac{\hat{\theta}_1 - \hat{\theta}_2}{\sigma_2 - \sigma_1}\bigg\rvert
  \geq -z_{\frac{\sqrt{w_1w_2}}{w_1+w_2}\sqrt{\alpha}}.
\end{align*}

\subsubsection{Bounds on Edgington's CI width in the well-separated case}

We start with the case that $p_1(\mu)$ and $p_2(\mu)$ are well-separated. Note
that $p_1(\mu)$ and $p_2(\mu)$ cannot be well-separated at $\alpha < 1$ when
$\hat{\theta}_1 = \hat{\theta}_2$ because then $p_{W}(\mu^\ast) = 0.5$.

We first assume that $\hat{\theta}_1 < \hat{\theta}_2$. As mentioned before,
under the assumption $\alpha < w_1/w_2$, the expression $p_{W}(\mu^\ast) \leq
\alpha/2 < w_1/(2w_2)$ is equivalent to the expression $w_1p_1(\mu^\ast) +
w_2p_2(\mu^\ast) \leq \sqrt{w_1w_2}\sqrt{\alpha}$, which is equivalent to the
expression $(\hat{\theta}_1-\hat{\theta}_2)/(\sigma_2 - \sigma_1 )\leq
z_{\frac{\sqrt{w_1w_2}}{w_1+w_2}\sqrt{\alpha}}$. Under the assumption $\alpha <
w_1/w_2$ (and, as stated in the main text, assuming without loss of generality
that $0 < w_1 \leq w_2$), we also have
$z_{\frac{\sqrt{w_1w_2}}{w_1+w_2}\sqrt{\alpha}} < z_{1/2} = 0$. Since we assume
that $\hat{\theta}_1 < \hat{\theta}_2$, these results together imply that
$\sigma_2 > \sigma_1$. This means that for all $\mu \geq \mu^\ast$, it holds
that $\mu (1/\sigma_1 - 1/\sigma_2) \geq \hat{\theta}_1/\sigma_1 -
\hat{\theta}_2/\sigma_2$, which is equivalent to $p_1(\mu) \geq p_2(\mu)$. Since
$w_1p_1(\mu) + w_2p_2(\mu)$ is strictly monotonically increasing in $\mu$, it
then follows that $\hat{\mu}_W(1-\alpha/2) > \hat{\mu}_W(\alpha/2) \geq
\mu^\ast$, where $\hat{\mu}_W(\gamma)$ is defined as the function such that
$p_W(\hat{\mu}_W(\gamma)) = \gamma$.
The weighted sum of the \textit{p}-values $s_w(\mu)$ can hence be bounded by
\begin{align*}
  \begin{cases}
    w_1p_1 (\mu) < w_1p_1(\mu) + w_2p_2(\mu) \leq (w_1+w_2) \ p_1(\mu) & \text{if $  \mu^\ast\leq \mu \leq \hat{\mu}_W(\frac{w_1}{2w_2})$,}\\
    (w_1+w_2) \ p_2(\mu) < w_1p_1(\mu) + w_2p_2(\mu) < w_1+ w_2p_2(\mu) & \text{if $\mu \geq\hat{\mu}_W(1-\frac{w_1}{2w_2})$,}
  \end{cases}
\end{align*}
which essentially means that in the region where $p_1(\mu)$ is increasing and
$p_2(\mu)$ remains close to 0 we set $p_2(\mu)$ to 0 (lower bound of the first
case), while in the region where $p_2(\mu)$ is increasing and $p_1(\mu)$ is
close to 1 we set $p_1(\mu)$ to 1 (upper bound of the second case). For
$p_W(\mu) = \alpha/2$ (i.e., $w_1p_1(\mu) +w_2p_2(\mu) = \sqrt{w_1w_2\alpha}$
obtained by setting $\mu = \hat{\mu}_W(\frac{\alpha}{2})$), we get
\begin{align*}
    w_1\Phi\left(\frac{\hat{\mu}_{W}(\alpha/2) - \hat{\theta}_1}{\sigma_1}\right) < \sqrt{w_1w_2}\sqrt{\alpha}  \leq  (w_1+w_2) \ \Phi\left(\frac{\hat{\mu}_{W}(\alpha/2) - \hat{\theta}_1}{\sigma_1}\right).
\end{align*}
Hence, the lower limit of the CI ($\hat{\mu}_{W}(\alpha/2)$) is bounded by
\begin{align*}
  \hat{\theta}_1 + \sigma_1 z_{\frac{\sqrt{w_1w_2}}{w_1+w_2}\sqrt{\alpha}}\leq \hat{\mu}_{W}(\alpha/2) < \hat{\theta}_1 + \sigma_1 z_{\sqrt{\frac{w_2}{w_1}}\sqrt{\alpha}}.
\end{align*}
Similarly, for $p_W(\mu) = 1 - \alpha/2$ (i.e., $w_1p_1(\mu) +w_2p_2(\mu) = (w_1 + w_2) -
\sqrt{w_1w_2\alpha}$ obtained by setting $\mu = \hat{\mu}_W(1 -
\frac{\alpha}{2})$), the upper limit of the CI ($\hat{\mu}_{W}(1-\alpha/2)$) is
bounded by
\begin{align*}
  \hat{\theta}_2 - \sigma_2 z_{\sqrt{\frac{w_1}{w_2}}\sqrt{\alpha}} < \hat{\mu}_{W}(1-\alpha/2) < \hat{\theta}_2 - \sigma_2 z_{\frac{\sqrt{w_1w_2}}{w_1+w_2}\sqrt{\alpha}}.
\end{align*}
Assuming $\alpha < w_1/(4w_2)$, we have $\hat{\theta}_1 + \sigma_1
z_{\sqrt{\frac{w_2}{w_1}}\sqrt{\alpha}}<\hat{\theta}_1<\hat{\theta}_2 <
\hat{\theta}_2 - \sigma_2 z_{\sqrt{\frac{w_1}{w_2}}\sqrt{\alpha}}$, which
ensures that the upper bound of the lower limit is strictly smaller than the
lower bound of the upper limit of the CI. The width of a $(1-\alpha) \times
100\%$ CI is thus bounded by
\begin{align*}
  \left(\hat{\theta}_2 -\hat{\theta}_1\right) - (\sigma_1z_{\sqrt{\frac{w_2}{w_1}}\sqrt{\alpha}}+\sigma_2 z_{\sqrt{\frac{w_1}{w_2}}\sqrt{\alpha}}) < \text{width(CI)} < \left(\hat{\theta}_2 -\hat{\theta}_1\right) - (\sigma_1+\sigma_2) z_{\frac{\sqrt{w_1w_2}}{w_1+w_2}\sqrt{\alpha}}.
\end{align*}
If $p_{W}(\mu^\ast) \geq 1-\alpha/2 > 1- w_1/(2w_2)$, then we have $\sigma_2 <
\sigma_1$ under the assumption $\hat{\theta}_1 < \hat{\theta}_2$. Then, it
follows that $p_1(\mu) \geq p_2(\mu)$ for all $\mu \leq \mu^\ast$, and we can
use the same arguments as above to obtain the same bounds for the CI width.

If $\hat{\theta}_1 > \hat{\theta}_2$, by similar arguments, we have the bounds
as
\begin{align*}
  \left(\hat{\theta}_1 -\hat{\theta}_2\right) - (\sigma_1z_{\sqrt{\frac{w_2}{w_1}}\sqrt{\alpha}}+\sigma_2 z_{\sqrt{\frac{w_1}{w_2}}\sqrt{\alpha}}) < \text{width(CI)} < \left(\hat{\theta}_1 -\hat{\theta}_2\right) - (\sigma_1+\sigma_2) z_{\frac{\sqrt{w_1w_2}}{w_1+w_2}\sqrt{\alpha}}.
\end{align*}
The case $\hat{\theta}_1 = \hat{\theta}_2$ is not possible, as we assume that
$p_1(\mu)$ and $p_2(\mu)$ are well-separated.

Thus, if $p_1(\mu)$ and $p_2(\mu)$ are well-separated at level $\alpha$ with
$\alpha < w_1/(4w_2)$, we have
\begin{align*}
  \lvert \hat{\theta}_2 -\hat{\theta}_1\rvert -  (\sigma_1z_{\sqrt{\frac{w_2}{w_1}}\sqrt{\alpha}}+\sigma_2 z_{\sqrt{\frac{w_1}{w_2}}\sqrt{\alpha}}) < \text{width(CI)} < \lvert \hat{\theta}_2 -\hat{\theta}_1\rvert - (\sigma_1+\sigma_2) z_{\frac{\sqrt{w_1w_2}}{w_1+w_2}\sqrt{\alpha}}.
\end{align*}

\subsubsection{Bounds on Edgington's CI width in the not well-separated case}

We continue with the case that $p_1(\mu)$ and $p_2(\mu)$ are not well-separated.
If $\sigma_1 < \sigma_2$, then for all $\mu > \mu^\ast$ we have $p_1(\mu) >
p_2(\mu)$ and for all $\mu < \mu^\ast$ we have $p_1(\mu) <
p_2(\mu)$\footnote{This follows from the fact that if $\sigma_1 < \sigma_2$,
then $1/\sigma_1 - 1/\sigma_2 > 0$. Then, in the case $\mu > \mu^\ast$, we have
$\mu > \frac{\hat{\theta}_1/\sigma_1- \hat{\theta}_2/\sigma_2}{1/\sigma_1 -
  1/\sigma_2}$ and because $1/\sigma_1 - 1/\sigma_2 > 0$, we have $(1/\sigma_1 -
1/\sigma_2 )\mu > \hat{\theta}_1/\sigma_1- \hat{\theta}_2/\sigma_2$, which is
equivalent to $\frac{\mu - \hat{\theta}_1}{\sigma_1} > \frac{\mu -
  \hat{\theta}_2}{\sigma_2} $. Since $\Phi(\cdot)$ is strictly monotonically
increasing, we have $p_1(\mu) = \Phi(\frac{\mu - \hat{\theta}_1}{\sigma_1}) >
\Phi( \frac{\mu - \hat{\theta}_2}{\sigma_2} ) = p_2(\mu)$.}. Therefore, the
weighted sum of \textit{p}-value functions is bounded by
\begin{align*}
  \begin{cases}
    w_2p_2 (\mu) < w_1p_1(\mu) + w_2p_2(\mu) \leq (w_1+w_2) \ p_2(\mu) & \text{if $ \mu < \mu^\ast$,}\\
    (w_1+w_2) \ p_2(\mu) < w_1p_1(\mu) + w_2p_2(\mu) <w_1+ w_2p_2(\mu) & \text{if $\mu > \mu^\ast$.}
  \end{cases}
\end{align*}
Since $p_1(\mu)$ and $p_2(\mu)$ are not well-separated ($\alpha/2 <
p_W(\mu^\ast) < 1 - \alpha/2 $) and $p_{W}(\mu)$ is monotonically increasing in
$\mu$, we then have $\hat{\mu}_{W}(1-\alpha/2) > \mu^\ast >
\hat{\mu}_{W}(\alpha/2)$. Combined with the above bounds, this means that the
lower limit of the CI ($\hat{\mu}_{W}(\alpha/2)$) is bounded by
\begin{align*}
  \hat{\theta}_2 + \sigma_2 z_{\frac{\sqrt{w_1w_2}}{w_1+w_2}\sqrt{\alpha}}< \hat{\mu}_{W}(\alpha/2) < \hat{\theta}_2 + \sigma_2 z_{\sqrt{\frac{w_1}{w_2}}\sqrt{\alpha}}
\end{align*}
and the upper limit of the CI ($\hat{\mu}_{W}(1-\alpha/2)$) is bounded by
\begin{align*}
  \hat{\theta}_2 - \sigma_2 z_{\sqrt{\frac{w_1}{w_2}}\sqrt{\alpha}} < \hat{\mu}_{W}(1-\alpha/2) < \hat{\theta}_2 - \sigma_2 z_{\frac{\sqrt{w_1w_2}}{w_1+w_2}\sqrt{\alpha}}.
\end{align*}
Assuming again $\alpha < w_1/(4w_2)$ to ensure that the upper bound of the lower
CI limit is strictly smaller than the lower bound of the upper CI limit, we have
$\hat{\theta}_2 + \sigma_2 z_{\sqrt{\frac{w_1}{w_2}}\sqrt{\alpha}}<
\hat{\theta}_2 - \sigma_2 z_{\sqrt{\frac{w_1}{w_2}}\sqrt{\alpha}} $. Thus, the
width of a $(1-\alpha) \times 100\%$ CI is bounded by
\begin{align*}
  - 2\ \sigma_2 z_{\sqrt{\frac{w_1}{w_2}}\sqrt{\alpha}} < \text{width(CI)} <  - 2\ \sigma_2 z_{\frac{\sqrt{w_1w_2}}{w_1+w_2}\sqrt{\alpha}}.
\end{align*}
If $\sigma_1 > \sigma_2$\footnote{If $\sigma_1 = \sigma_2$ and $\hat{\theta}_1
\neq \hat{\theta}_2$, then $p_1(\mu)$ and $p_2(\mu)$ are well-separated. If
$\sigma_1 = \sigma_2$ and $\hat{\theta}_1 = \hat{\theta}_2$, $p_1(\mu) =
p_2(\mu)$ and the CI is available in closed-form and narrower than fixed-effect
meta-analysis, see Pawel et al. \citep[Section~3.3.1]{Pawel2025}.}, then we have
by similar arguments
\begin{align*}
  - 2\ \sigma_1 z_{\sqrt{\frac{w_2}{w_1}}\sqrt{\alpha}} < \text{width(CI)} <  - 2\ \sigma_1 z_{\frac{\sqrt{w_1w_2}}{w_1+w_2}\sqrt{\alpha}}.
\end{align*}
Thus, if $p_1(\mu)$ and $p_2(\mu)$ are not well-separated at level $\alpha$ with
$\alpha < w_1/(4w_2)$, we have
\begin{align*}
   &\left(- 2\, \sigma_2 z_{\sqrt{\frac{w_1}{w_2}}\sqrt{\alpha}}\right)
   \cdot \mathbb{1}(\sigma_2 > \sigma_1)
   + \left(- 2\, \sigma_1 z_{\sqrt{\frac{w_2}{w_1}}\sqrt{\alpha}}\right)
   \cdot \mathbb{1}(\sigma_1 > \sigma_2) < \\
   &\text{width(CI)}
   < - 2\max\left\{\sigma_1, \,\sigma_2\right\}\,
   z_{\frac{\sqrt{w_1w_2}}{w_1+w_2}\sqrt{\alpha}}.
\end{align*}

\subsection{Conditions for narrower Edgington's CI}

The width of the $(1-\alpha) \times 100\%$ CI based on inverse-variance
fixed-effect meta-analysis is given by
\begin{align*}
  \text{width}(\text{CI}_{\text{FE}})
  = (2 \, z_{1 - \alpha/2}) \big / \sqrt{1/\sigma_1^2 + 1/\sigma_2^2}.
\end{align*}
This expression can be used to find a sufficient but not necessary condition for
Edgington's CI to be narrower (again under the general assumptions $\alpha <
w_1/(4w_2)$ and $\sigma_1 \neq \sigma_2$; see
Section~\ref{sec:bounds-edg-width} where also the case $\sigma_1 = \sigma_2$
is explained). Specifically, if the upper bound on the width of Edgington's CI
is smaller than the width of the CI based on inverse-variance fixed-effect
meta-analysis, then Edgington's method will have a narrower CI.
\begin{itemize}
    \item If $p_1(\mu)$ and $p_2(\mu)$ are well-separated,
      the condition is that
\begin{align*}
    \text{width}(\text{CI}_{\text{E}}) < \lvert \hat{\theta}_2 -\hat{\theta}_1\rvert - (\sigma_1+\sigma_2) z_{\frac{\sqrt{w_1w_2}}{w_1+w_2}\sqrt{\alpha}} < \frac{2 \, z_{1 - \alpha/2}}  {\sqrt{1/\sigma_1^2 + 1/\sigma_2^2}} = \text{width}(\text{CI}_{\text{FE}}),
\end{align*}
which can be summarized as
\begin{align}
      \label{eq:cond-well-sep}
   -z_{\frac{\sqrt{w_1w_2}}{w_1+w_2}\sqrt{\alpha}}\ \lvert \sigma_2 - \sigma_1 \rvert \leq \lvert \hat{\theta}_1 -\hat{\theta}_2\rvert  < \frac{2 \, z_{1 - \alpha/2}}  {\sqrt{1/\sigma_1^2 + 1/\sigma_2^2}} +(\sigma_1+\sigma_2) z_{\frac{\sqrt{w_1w_2}}{w_1+w_2}\sqrt{\alpha}}.
\end{align}
 \item If $p_1(\mu)$ and $p_2(\mu)$ are not well-separated,
   the condition is that
 \begin{align*}
        \text{width}(\text{CI}_{\text{E}}) <  - 2\max{\left\{\sigma_1, \,\sigma_2\right\}} \ z_{\frac{\sqrt{w_1w_2}}{w_1+w_2}\sqrt{\alpha}} < \frac{2 \, z_{1 - \alpha/2}}  {\sqrt{1/\sigma_1^2 + 1/\sigma_2^2}} =  \text{width}(\text{CI}_{\text{FE}}),
    \end{align*}
    which can be summarized as
    \begin{align}
      \label{eq:cond-no-well-sep}
         \lvert \hat{\theta}_1 - \hat{\theta}_2\rvert < -z_{\frac{\sqrt{w_1w_2}}{w_1+w_2}\sqrt{\alpha}} \ \lvert \sigma_2 - \sigma_1 \rvert \quad \text{and} \quad\frac{\max{\{\sigma_1,\,\sigma_2\}}}{\min{\{\sigma_1,\,\sigma_2\}}} < \sqrt{\left(\frac{z_{\alpha/2}}{z_{\frac{\sqrt{w_1w_2}}{w_1+w_2}\sqrt{\alpha}}}\right)^2-1}.
    \end{align}
\end{itemize}
Note that in~\eqref{eq:cond-well-sep}, the lower bound is only smaller than the
upper bound (and thus the range for $\lvert \hat{\theta}_1
-\hat{\theta}_2\rvert$ to fall into and satisfy the condition is only nonempty)
if
\begin{align}
  \label{eq:nonempty-cond}
     -z_{\frac{\sqrt{w_1w_2}}{w_1+w_2}\sqrt{\alpha}}\ \lvert \sigma_2 - \sigma_1 \rvert<-\frac{2 \, z_{ \alpha/2}}  {\sqrt{1/\sigma_1^2 + 1/\sigma_2^2}} +(\sigma_1+\sigma_2) z_{\frac{\sqrt{w_1w_2}}{w_1+w_2}\sqrt{\alpha}}.
\end{align}
By writing $\lvert \sigma_2 - \sigma_1 \rvert = \max{\{ \sigma_1,\, \sigma_2\}}
- \min{\{ \sigma_1,\, \sigma_2\}}$ and $\sigma_1+\sigma_2 = \max{\{ \sigma_1,\,
  \sigma_2\}} + \min{\{ \sigma_1,\, \sigma_2\}}$, the above
inequality~\eqref{eq:nonempty-cond} is equivalent to
$\left(\frac{\max{\{\sigma_1,\,\sigma_2\}}}{\min{\{\sigma_1,\,\sigma_2\}}}\right)^2
<
\left(\frac{z_{\alpha/2}}{z_{\frac{\sqrt{w_1w_2}}{w_1+w_2}\sqrt{\alpha}}}\right)^2-1$,
the second condition that is part of~\eqref{eq:cond-no-well-sep}.

%
The two cases partition the $\lvert \hat{\theta}_1 - \hat{\theta}_2 \rvert$ axis
at $-z_{\frac{\sqrt{w_1w_2}}{w_1+w_2}\sqrt{\alpha}}\, \lvert \sigma_2 - \sigma_1
\rvert$: condition~\eqref{eq:cond-no-well-sep} covers values below this point and
condition~\eqref{eq:cond-well-sep} values from this point up to the upper bound.
Since the nonemptiness condition~\eqref{eq:nonempty-cond} of the well-separated
range is identical to the variance-ratio condition
in~\eqref{eq:cond-no-well-sep}, both cases require it, and under it the two
ranges meet at their shared endpoint and combine into the single interval
\begin{align*}
  \lvert \hat{\theta}_1 - \hat{\theta}_2 \rvert
  < \frac{2 \, z_{1 - \alpha/2}}{\sqrt{1/\sigma_1^2 + 1/\sigma_2^2}}
  + (\sigma_1 + \sigma_2)\, z_{\frac{\sqrt{w_1w_2}}{w_1+w_2}\sqrt{\alpha}}.
\end{align*}
This thus allows us to combine the cases for
well-separated~\eqref{eq:cond-well-sep} and not
well-separated~\eqref{eq:cond-no-well-sep} \textit{p}-value functions into a
single set of conditions. Assuming that $\alpha < w_1/(4w_2)$ and $\sigma_1 \neq
\sigma_2$ (see Section~\ref{sec:bounds-edg-width} for the case $\sigma_1 =
\sigma_2$), Edgington's method will yield a narrower CI if the following
conditions hold:
\begin{itemize}
  \item[] $\lvert \hat{\theta}_1 -\hat{\theta}_2\rvert  < \frac{2 \, z_{1 - \alpha/2}}  {\sqrt{1/\sigma_1^2 + 1/\sigma_2^2}} +(\sigma_1+\sigma_2) z_{\frac{\sqrt{w_1w_2}}{w_1+w_2}\sqrt{\alpha}}$ and
     $\frac{\max{\{\sigma_1,\,\sigma_2\}}}{\min{\{\sigma_1,\,\sigma_2\}}} < \sqrt{\left(\frac{z_{\alpha/2}}{z_{\frac{\sqrt{w_1w_2}}{w_1+w_2}\sqrt{\alpha}}}\right)^2-1}$.
\end{itemize}
These two conditions can then be transformed into conditions on $Q =
(\hat{\theta}_1 - \hat{\theta}_2)^2/(\sigma^2_1 + \sigma^2_2)$ and $r =
\sigma_1/\sigma_2$ shown in Section~\ref{sec:Edgingtonconditions}, by expressing
$\lvert \hat{\theta}_1 -\hat{\theta}_2\rvert = \sqrt{Q (\sigma^2_1 +
  \sigma^2_2)}$ in terms of $Q$ and similarly $\sigma_1 = r \sigma_2$ in terms
of $r$, followed by algebraic manipulations.

\subsection{Conditions for wider Edgington's CI}

In a similar way as for a narrower CI, it is possible to derive conditions for
Edgington's method producing a wider CI than fixed-effect meta-analysis.
Specifically, if the lower bound on the width of Edgington's CI is larger than
the width of the CI based on inverse-variance fixed-effect meta-analysis, then
the weighted (and unweighted) Edgington's method will have a wider CI. For the
unweighted case, assuming $\alpha < 1/4$ and $\sigma_1 \neq \sigma_2$ (see
Section~\ref{sec:bounds-edg-width} for the case $\sigma_1 = \sigma_2$), in
the well-separated case Edgington's CI is wider if $\lvert \hat{\theta}_2
-\hat{\theta}_1\rvert - (\sigma_1+\sigma_2) z_{\sqrt{\alpha}} >
\text{width}(\text{CI}_{\text{FE}})$ while in the not well-separated case it is
wider if $-2 z_{\sqrt{\alpha}} \max\{\sigma_1, \sigma_2\} >
\text{width}(\text{CI}_{\text{FE}})$.
These conditions can again be transformed into conditions on $Q$ and $r$:
\begin{enumerate}
    \item $Q > \bigg\{z_{\sqrt{\alpha}}\,\frac{r + 1}{\sqrt{r^2 + 1}} -
      z_{\alpha/2} \, \frac{2 \, r}{1 + r^2} \bigg\}^2$ and $Q \geq
      z_{\sqrt{\alpha/4}}^2 \,\frac{(1 - r)^2}{r^2 + 1}$,
  \item $\max\left\{r^2, 1/r^2\right\}
  > \left(\frac{z_{\alpha/2}}{z_{\sqrt{\alpha}}}\right)^2 - 1$ and $Q <
   z_{\sqrt{\alpha/4}}^2 \,\frac{(1 - r)^2}{r^2 + 1}$.
\end{enumerate}

From these, looser conditions that only depend on $Q$ or $r$ can be derived (for
a given $\alpha$). Note first that cases 1 and 2 are mutually exclusive, since
their second inequalities are complementary. A sufficient condition involving
only one variable must therefore guarantee that the relevant case holds across
the entire range of the other. For $Q$, this means exceeding the lower bound in
case 1 for every $r$; that bound is maximized at $r = 1$, giving $Q >
\left(\sqrt{2} z_{\sqrt{\alpha}} - z_{\alpha/2}\right)^2$. For $r$, the
inequality in case 2 does not involve $Q$, so it serves directly as the bound,
giving $\max\left\{r, 1/r\right\} >
\sqrt{\left(\frac{z_{\alpha/2}}{z_{\sqrt{\alpha}}}\right)^2 - 1}$. Edgington's
CI is thus wider than the fixed-effect meta-analysis CI whenever either holds.
For $\alpha = 0.05$, these translate into $\max\left\{r, 1/r\right\} >
2.38$ and $Q > 0.78$ (rounded), which
will also be shown graphically in the following subsection.

The same derivation applies to the weighted method, but the expressions become
more complicated because the lower limit of its CI depends on whether $\sigma_1
> \sigma_2$ or $\sigma_1 < \sigma_2$.

\subsection{Empirical evaluation of narrower/wider conditions}

\begin{figure}[!htb]
\begin{knitrout}
\definecolor{shadecolor}{rgb}{0.969, 0.969, 0.969}\color{fgcolor}

{\centering \includegraphics[width=\maxwidth]{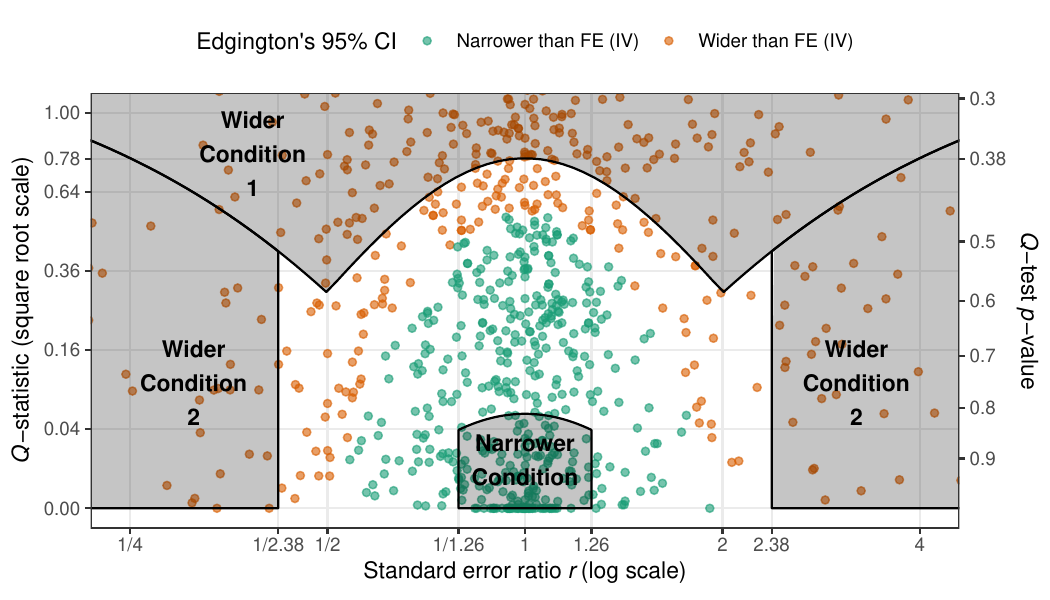} 

}

\end{knitrout}
\caption{$Q$-statistic and standard error ratio $r = \sigma_1/\sigma_2$ of IQWiG
  two-study meta-analyses, along with the sufficient condition regions for
  Edgington's method (unweighted) giving narrower/wider 95\% CIs than
  inverse-variance fixed-effect meta-analysis.}
\label{fig:iqwigCIconditions}
\end{figure}

Figure~\ref{fig:iqwigCIconditions} shows the $Q$-statistic versus the standard
error ratio $r = \sigma_1/\sigma_2$ for two-study meta-analyses from the IQWiG
database. Meta-analyses where Edgington's (unweighted) method yields a narrower
CI than inverse-variance fixed-effect meta-analysis are colored in green while
meta-analyses with wider CIs are colored in orange.

We can see that meta-analyses with narrower CIs include studies with standard
errors that differ by no more than a factor of two ($1/2 < r < 2$) and have a
relatively small $Q$-statistic ($Q < 0.6$). Conversely, meta-analyses with wider
CIs show more unequal standard errors or larger $Q$-statistics.

The plot also shows the regions where the sufficient conditions for a narrower
or wider CI apply. In $(154 +
663)/1226 = 66.6\%$ of
meta-analyses, either the narrower or the wider CI condition is fulfilled. In
the remaining cases, no condition is satisfied, demonstrating that these are
only sufficient but not necessary conditions. Finally,
$614/1226 $ (50.1\%)
meta-analyses satisfy $\max\left\{r, 1/r\right\} > 2.38$ or
$Q > 0.78$ (and show a wider CI than fixed-effect
meta-analysis), consistent with these being looser but also sufficient conditions
for a wider CI.

\section{Additional CI width comparisons}
\label{app:CIwidthresults}
\setcounter{figure}{0} \setcounter{table}{0}

\begin{figure}[!htb]
\begin{knitrout}
\definecolor{shadecolor}{rgb}{0.969, 0.969, 0.969}\color{fgcolor}

{\centering \includegraphics[width=\maxwidth]{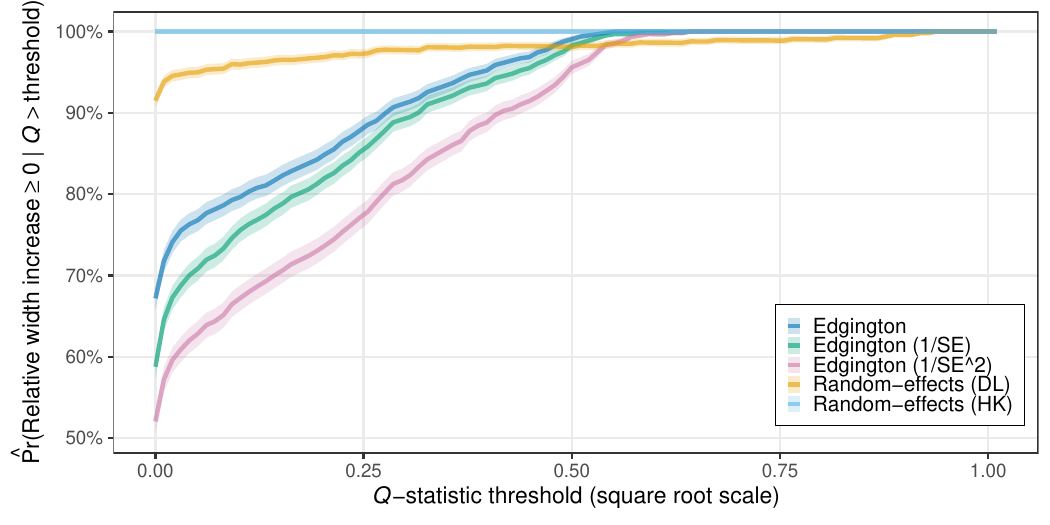} 

}

\end{knitrout}
\caption{The proportion of IQWiG two-study meta-analyses with 95\% CI width
  equal or increased relative to fixed-effect meta-analysis given that $Q$ is
  greater than the threshold on the x-axis. Error bands indicate point-wise
  standard error ranges.}
\label{fig:iqwigCIcomparisonApp}
\end{figure}

Figure~\ref{fig:iqwigCIcomparisonApp} illustrates the same patterns as in
Figure~\ref{fig:iqwigmethodcomparison}b but at an aggregated level. Shown is the
proportion of CIs with equal or increased width relative to fixed-effect
meta-analysis among meta-analyses with a $Q$-statistic exceeding a certain
threshold (x-axis).

HK random-effects meta-analysis shows equal or increased CI width compared to
fixed-effect meta-analysis in all meta-analyses and at all $Q$ thresholds. DL
random-effects meta-analysis behaves similarly, with equal or increased CI width
in more than 90\% of meta-analyses even at $Q = 0$. The few exceptions arise
when the MH method is used for fixed-effect meta-analysis, which can produce
wider CIs than DL random-effects meta-analysis. DL random-effects meta-analysis
plateaus at 100\% at $Q = 1$, as expected since the DL/PM/REML estimator yields
a positive heterogeneity estimate only when $Q > 1$.

The Edgington variants show a different pattern. Since their CIs can also be
narrower than fixed-effect meta-analysis, the proportion of meta-analyses with
equal or increased CI width starts well below 100\% but rises with increasing
$Q$. The unweighted variant increases fastest, plateauing at 100\% shortly after
$Q = 0.5$, followed closely by the inverse-standard-error weighted variant,
which shows a slightly slower but otherwise similar pattern. The
inverse-variance weighted variant starts lowest and plateaus at 100\% around $Q
= 0.6$.

\section{Significance disagreement}
\label{app:significance}
\setcounter{figure}{0}
\setcounter{table}{0}

Table~\ref{tab:conflict} lists all meta-analyses where fixed-effect
meta-analysis and at least one Edgington's method give different conclusions
regarding statistical significance.

\begingroup\fontsize{6.5}{8.5}\selectfont

\begin{longtable}[t]{llrlllll}
\caption{\label{tab:conflict}Point estimates and 95\% CIs from two-study IQWiG meta-analyses with statistical significance disagreement ($\alpha = 0.05$) between fixed-effect meta-analysis and at least one Edgington's method. The meta-analyses are ordered by  $Q$-statistic and colored according to statistical significance (blue: non-significant; orange: significant).}\\
\toprule
\textbf{Study 1} & \textbf{Study 2} & \textbf{$\boldsymbol{Q}$} & \textbf{Fixed-effect} & \textbf{Edgington} & \textbf{Edgington (1/SE)} & \textbf{Edgington (1/$\text{SE}^2$)} & \textbf{Random-effects (DL)}\\
\midrule
\endfirsthead
\caption[]{Point estimates and 95\% CIs from two-study IQWiG meta-analyses with statistical significance disagreement ($\alpha = 0.05$) between fixed-effect meta-analysis and at least one Edgington's method. The meta-analyses are ordered by  $Q$-statistic and colored according to statistical significance (blue: non-significant; orange: significant). \textit{(continued)}}\\
\toprule
\textbf{Study 1} & \textbf{Study 2} & \textbf{$\boldsymbol{Q}$} & \textbf{Fixed-effect} & \textbf{Edgington} & \textbf{Edgington (1/SE)} & \textbf{Edgington (1/$\text{SE}^2$)} & \textbf{Random-effects (DL)}\\
\midrule
\endhead

\endfoot
\bottomrule
\endlastfoot
\cellcolor{Gray!10}{-0.4 (-0.7, -0.1)} & \cellcolor{Gray!10}{8.8 (7.2, 10.4)} & \cellcolor{Gray!10}{122.3} & \cellcolor{Gray!10}{\color[HTML]{0072B2}-0.1 (-0.4, 0.2)} & \cellcolor{Gray!10}{\color[HTML]{0072B2}1.8 (-0.5, 9.4)} & \cellcolor{Gray!10}{\color[HTML]{0072B2}-0.4 (-0.6, 8.8)} & \cellcolor{Gray!10}{\color[HTML]{D55E00}-0.4 (-0.7, -0.0)} & \cellcolor{Gray!10}{\color[HTML]{0072B2}4.2 (-4.8, 13.2)}\\
0.1 (-0.2, 0.4) & -2.1 (-2.5, -1.7) & 62.9 & \color[HTML]{D55E00}-0.7 (-0.9, -0.4) & \color[HTML]{0072B2}-0.8 (-2.3, 0.2) & \color[HTML]{0072B2}-0.1 (-2.2, 0.2) & \color[HTML]{0072B2}-0.0 (-2.2, 0.3) & \color[HTML]{0072B2}-1.0 (-3.2, 1.2)\\
\cellcolor{Gray!10}{-2.3 (-2.7, -1.8)} & \cellcolor{Gray!10}{0.0 (-0.5, 0.5)} & \cellcolor{Gray!10}{43.7} & \cellcolor{Gray!10}{\color[HTML]{D55E00}-1.3 (-1.6, -0.9)} & \cellcolor{Gray!10}{\color[HTML]{0072B2}-1.2 (-2.5, 0.2)} & \cellcolor{Gray!10}{\color[HTML]{0072B2}-1.9 (-2.5, 0.2)} & \cellcolor{Gray!10}{\color[HTML]{0072B2}-2.0 (-2.5, 0.2)} & \cellcolor{Gray!10}{\color[HTML]{0072B2}-1.1 (-3.4, 1.1)}\\
0.1 (-0.2, 0.4) & -1.4 (-1.7, -1.1) & 41.9 & \color[HTML]{D55E00}-0.7 (-0.9, -0.4) & \color[HTML]{0072B2}-0.7 (-1.5, 0.2) & \color[HTML]{0072B2}-1.0 (-1.5, 0.2) & \color[HTML]{0072B2}-1.1 (-1.5, 0.2) & \color[HTML]{0072B2}-0.7 (-2.1, 0.8)\\
\cellcolor{Gray!10}{4.0 (-5.0, 13.0)} & \cellcolor{Gray!10}{-25.0 (-29.9, -20.1)} & \cellcolor{Gray!10}{31.0} & \cellcolor{Gray!10}{\color[HTML]{D55E00}-18.3 (-22.6, -14.0)} & \cellcolor{Gray!10}{\color[HTML]{0072B2}-14.8 (-26.9, 7.5)} & \cellcolor{Gray!10}{\color[HTML]{0072B2}-23.1 (-27.4, 6.4)} & \cellcolor{Gray!10}{\color[HTML]{0072B2}-24.0 (-27.9, 5.0)} & \cellcolor{Gray!10}{\color[HTML]{0072B2}-10.8 (-39.2, 17.7)}\\
-1.3 (-1.7, -1.0) & 0.1 (-0.3, 0.5) & 27.6 & \color[HTML]{D55E00}-0.6 (-0.9, -0.3) & \color[HTML]{0072B2}-0.6 (-1.5, 0.3) & \color[HTML]{0072B2}-0.7 (-1.5, 0.3) & \color[HTML]{0072B2}-0.7 (-1.5, 0.3) & \color[HTML]{0072B2}-0.6 (-2.0, 0.8)\\
\cellcolor{Gray!10}{-1.3 (-1.7, -0.9)} & \cellcolor{Gray!10}{0.3 (-0.2, 0.8)} & \cellcolor{Gray!10}{25.6} & \cellcolor{Gray!10}{\color[HTML]{D55E00}-0.6 (-0.9, -0.3)} & \cellcolor{Gray!10}{\color[HTML]{0072B2}-0.6 (-1.4, 0.5)} & \cellcolor{Gray!10}{\color[HTML]{0072B2}-1.0 (-1.5, 0.5)} & \cellcolor{Gray!10}{\color[HTML]{0072B2}-1.1 (-1.5, 0.4)} & \cellcolor{Gray!10}{\color[HTML]{0072B2}-0.5 (-2.1, 1.0)}\\
0.0 (-0.5, 0.5) & -1.6 (-2.1, -1.1) & 20.4 & \color[HTML]{D55E00}-0.9 (-1.2, -0.5) & \color[HTML]{0072B2}-0.8 (-1.8, 0.2) & \color[HTML]{0072B2}-1.2 (-1.8, 0.2) & \color[HTML]{0072B2}-1.3 (-1.8, 0.2) & \color[HTML]{0072B2}-0.8 (-2.4, 0.8)\\
\cellcolor{Gray!10}{-0.9 (-1.3, -0.6)} & \cellcolor{Gray!10}{0.2 (-0.2, 0.6)} & \cellcolor{Gray!10}{17.1} & \cellcolor{Gray!10}{\color[HTML]{D55E00}-0.5 (-0.8, -0.3)} & \cellcolor{Gray!10}{\color[HTML]{0072B2}-0.4 (-1.1, 0.3)} & \cellcolor{Gray!10}{\color[HTML]{0072B2}-0.7 (-1.1, 0.3)} & \cellcolor{Gray!10}{\color[HTML]{0072B2}-0.8 (-1.1, 0.3)} & \cellcolor{Gray!10}{\color[HTML]{0072B2}-0.4 (-1.5, 0.7)}\\
-1.6 (-1.7, -1.5) & -0.1 (-1.0, 0.8) & 11.7 & \color[HTML]{D55E00}-1.6 (-1.7, -1.5) & \color[HTML]{0072B2}-1.5 (-1.6, 0.2) & \color[HTML]{D55E00}-1.6 (-1.7, -0.5) & \color[HTML]{D55E00}-1.6 (-1.7, -1.5) & \color[HTML]{0072B2}-0.9 (-2.4, 0.6)\\
\cellcolor{Gray!10}{-13.3 (-18.7, -7.9)} & \cellcolor{Gray!10}{-1.0 (-5.8, 3.8)} & \cellcolor{Gray!10}{11.1} & \cellcolor{Gray!10}{\color[HTML]{D55E00}-6.3 (-9.9, -2.8)} & \cellcolor{Gray!10}{\color[HTML]{0072B2}-6.7 (-15.4, 0.8)} & \cellcolor{Gray!10}{\color[HTML]{0072B2}-4.7 (-15.3, 1.0)} & \cellcolor{Gray!10}{\color[HTML]{0072B2}-3.9 (-15.1, 1.1)} & \cellcolor{Gray!10}{\color[HTML]{0072B2}-7.1 (-19.1, 5.0)}\\
-2.5 (-8.1, 3.1) & 9.4 (4.4, 14.4) & 9.6 & \color[HTML]{D55E00}4.2 (0.4, 7.9) & \color[HTML]{0072B2}3.8 (-4.7, 11.3) & \color[HTML]{0072B2}5.4 (-4.6, 11.5) & \color[HTML]{0072B2}6.3 (-4.4, 11.6) & \color[HTML]{0072B2}3.5 (-8.1, 15.2)\\
\cellcolor{Gray!10}{5.5 (2.5, 8.5)} & \cellcolor{Gray!10}{0.5 (-1.2, 2.1)} & \cellcolor{Gray!10}{8.4} & \cellcolor{Gray!10}{\color[HTML]{D55E00}2.4 (1.3, 3.4)} & \cellcolor{Gray!10}{\color[HTML]{0072B2}2.3 (-0.2, 6.7)} & \cellcolor{Gray!10}{\color[HTML]{0072B2}1.1 (-0.3, 6.3)} & \cellcolor{Gray!10}{\color[HTML]{0072B2}0.8 (-0.5, 5.9)} & \cellcolor{Gray!10}{\color[HTML]{0072B2}2.8 (-2.1, 7.7)}\\
-0.1 (-0.6, 0.4) & -1.3 (-2.0, -0.7) & 8.3 & \color[HTML]{D55E00}-0.6 (-1.0, -0.1) & \color[HTML]{0072B2}-0.6 (-1.6, 0.1) & \color[HTML]{0072B2}-0.4 (-1.6, 0.1) & \color[HTML]{0072B2}-0.3 (-1.5, 0.2) & \color[HTML]{0072B2}-0.7 (-1.9, 0.5)\\
\cellcolor{Gray!10}{-0.3 (-2.1, 1.5)} & \cellcolor{Gray!10}{-3.2 (-4.3, -2.2)} & \cellcolor{Gray!10}{7.6} & \cellcolor{Gray!10}{\color[HTML]{D55E00}-2.6 (-3.4, -1.8)} & \cellcolor{Gray!10}{\color[HTML]{0072B2}-2.1 (-3.6, 0.4)} & \cellcolor{Gray!10}{\color[HTML]{0072B2}-2.8 (-3.7, 0.2)} & \cellcolor{Gray!10}{\color[HTML]{0072B2}-3.0 (-3.8, 0.0)} & \cellcolor{Gray!10}{\color[HTML]{0072B2}-1.8 (-4.7, 1.0)}\\
-0.1 (-0.2, 0.0) & -0.7 (-1.2, -0.3) & 7.2 & \color[HTML]{D55E00}-0.1 (-0.3, -0.0) & \color[HTML]{D55E00}-0.2 (-0.9, -0.0) & \color[HTML]{D55E00}-0.1 (-0.8, -0.0) & \color[HTML]{0072B2}-0.1 (-0.6, 0.0) & \color[HTML]{0072B2}-0.4 (-1.0, 0.3)\\
\cellcolor{Gray!10}{23.2 (13.0, 33.4)} & \cellcolor{Gray!10}{1.5 (-11.6, 14.6)} & \cellcolor{Gray!10}{6.6} & \cellcolor{Gray!10}{\color[HTML]{D55E00}14.9 (6.9, 23.0)} & \cellcolor{Gray!10}{\color[HTML]{0072B2}13.7 (-3.6, 27.2)} & \cellcolor{Gray!10}{\color[HTML]{0072B2}17.0 (-2.9, 27.6)} & \cellcolor{Gray!10}{\color[HTML]{0072B2}18.7 (-2.3, 28.1)} & \cellcolor{Gray!10}{\color[HTML]{0072B2}12.7 (-8.5, 34.0)}\\
0.2 (-0.3, 0.8) & -0.6 (-0.9, -0.2) & 6.5 & \color[HTML]{D55E00}-0.3 (-0.6, -0.0) & \color[HTML]{0072B2}-0.3 (-0.7, 0.4) & \color[HTML]{0072B2}-0.4 (-0.7, 0.4) & \color[HTML]{0072B2}-0.5 (-0.8, 0.3) & \color[HTML]{0072B2}-0.2 (-1.0, 0.6)\\
\cellcolor{Gray!10}{0.3 (-0.8, 1.3)} & \cellcolor{Gray!10}{-1.4 (-2.1, -0.6)} & \cellcolor{Gray!10}{6.5} & \cellcolor{Gray!10}{\color[HTML]{D55E00}-0.8 (-1.4, -0.2)} & \cellcolor{Gray!10}{\color[HTML]{0072B2}-0.7 (-1.7, 0.7)} & \cellcolor{Gray!10}{\color[HTML]{0072B2}-1.0 (-1.7, 0.6)} & \cellcolor{Gray!10}{\color[HTML]{0072B2}-1.1 (-1.8, 0.5)} & \cellcolor{Gray!10}{\color[HTML]{0072B2}-0.6 (-2.2, 1.1)}\\
-0.5 (-0.8, -0.2) & -0.0 (-0.3, 0.3) & 5.6 & \color[HTML]{D55E00}-0.3 (-0.5, -0.1) & \color[HTML]{0072B2}-0.3 (-0.6, 0.1) & \color[HTML]{0072B2}-0.3 (-0.6, 0.1) & \color[HTML]{0072B2}-0.3 (-0.6, 0.1) & \color[HTML]{0072B2}-0.3 (-0.7, 0.2)\\
\cellcolor{Gray!10}{-0.3 (-0.4, -0.1)} & \cellcolor{Gray!10}{-0.0 (-0.2, 0.1)} & \cellcolor{Gray!10}{5.5} & \cellcolor{Gray!10}{\color[HTML]{D55E00}-0.2 (-0.3, -0.1)} & \cellcolor{Gray!10}{\color[HTML]{0072B2}-0.2 (-0.3, 0.0)} & \cellcolor{Gray!10}{\color[HTML]{0072B2}-0.2 (-0.3, 0.0)} & \cellcolor{Gray!10}{\color[HTML]{0072B2}-0.2 (-0.3, 0.0)} & \cellcolor{Gray!10}{\color[HTML]{0072B2}-0.2 (-0.4, 0.1)}\\
-1.6 (-2.6, -0.7) & -0.2 (-0.9, 0.5) & 5.4 & \color[HTML]{D55E00}-0.8 (-1.4, -0.3) & \color[HTML]{0072B2}-0.8 (-2.0, 0.1) & \color[HTML]{0072B2}-0.6 (-1.9, 0.1) & \color[HTML]{0072B2}-0.5 (-1.9, 0.1) & \color[HTML]{0072B2}-0.9 (-2.3, 0.5)\\
\cellcolor{Gray!10}{-10.2 (-14.6, -5.8)} & \cellcolor{Gray!10}{-0.4 (-7.4, 6.6)} & \cellcolor{Gray!10}{5.3} & \cellcolor{Gray!10}{\color[HTML]{D55E00}-7.4 (-11.2, -3.7)} & \cellcolor{Gray!10}{\color[HTML]{0072B2}-6.4 (-11.9, 2.3)} & \cellcolor{Gray!10}{\color[HTML]{0072B2}-8.3 (-12.3, 1.7)} & \cellcolor{Gray!10}{\color[HTML]{0072B2}-9.1 (-12.6, 0.9)} & \cellcolor{Gray!10}{\color[HTML]{0072B2}-5.7 (-15.3, 3.9)}\\
0.0 (-0.3, 0.3) & -0.4 (-0.7, -0.2) & 5.1 & \color[HTML]{D55E00}-0.2 (-0.4, -0.0) & \color[HTML]{0072B2}-0.2 (-0.5, 0.1) & \color[HTML]{0072B2}-0.2 (-0.5, 0.1) & \color[HTML]{0072B2}-0.2 (-0.5, 0.1) & \color[HTML]{0072B2}-0.2 (-0.6, 0.2)\\
\cellcolor{Gray!10}{0.2 (-2.3, 2.7)} & \cellcolor{Gray!10}{-3.2 (-4.7, -1.7)} & \cellcolor{Gray!10}{5.0} & \cellcolor{Gray!10}{\color[HTML]{D55E00}-2.3 (-3.6, -1.0)} & \cellcolor{Gray!10}{\color[HTML]{0072B2}-1.9 (-3.8, 1.2)} & \cellcolor{Gray!10}{\color[HTML]{0072B2}-2.6 (-3.9, 0.9)} & \cellcolor{Gray!10}{\color[HTML]{0072B2}-2.8 (-4.1, 0.6)} & \cellcolor{Gray!10}{\color[HTML]{0072B2}-1.7 (-5.0, 1.7)}\\
-3.0 (-4.9, -1.1) & 0.0 (-1.9, 1.9) & 4.8 & \color[HTML]{D55E00}-1.5 (-2.8, -0.1) & \color[HTML]{0072B2}-1.5 (-3.7, 0.7) & \color[HTML]{0072B2}-1.4 (-3.7, 0.7) & \color[HTML]{0072B2}-1.4 (-3.7, 0.7) & \color[HTML]{0072B2}-1.5 (-4.4, 1.4)\\
\cellcolor{Gray!10}{-0.2 (-0.7, 0.2)} & \cellcolor{Gray!10}{-1.5 (-2.5, -0.5)} & \cellcolor{Gray!10}{4.8} & \cellcolor{Gray!10}{\color[HTML]{D55E00}-0.5 (-0.9, -0.0)} & \cellcolor{Gray!10}{\color[HTML]{D55E00}-0.6 (-1.9, -0.1)} & \cellcolor{Gray!10}{\color[HTML]{D55E00}-0.4 (-1.7, -0.0)} & \cellcolor{Gray!10}{\color[HTML]{0072B2}-0.3 (-1.5, 0.0)} & \cellcolor{Gray!10}{\color[HTML]{0072B2}-0.8 (-2.0, 0.4)}\\
2.7 (0.9, 4.5) & 0.1 (-1.5, 1.7) & 4.4 & \color[HTML]{D55E00}1.3 (0.2, 2.4) & \color[HTML]{0072B2}1.3 (-0.5, 3.4) & \color[HTML]{0072B2}1.1 (-0.6, 3.4) & \color[HTML]{0072B2}1.0 (-0.6, 3.3) & \color[HTML]{0072B2}1.4 (-1.2, 3.9)\\
\cellcolor{Gray!10}{-0.8 (-1.2, -0.4)} & \cellcolor{Gray!10}{-0.1 (-0.6, 0.4)} & \cellcolor{Gray!10}{4.3} & \cellcolor{Gray!10}{\color[HTML]{D55E00}-0.6 (-0.9, -0.3)} & \cellcolor{Gray!10}{\color[HTML]{0072B2}-0.5 (-1.0, 0.1)} & \cellcolor{Gray!10}{\color[HTML]{0072B2}-0.6 (-1.0, 0.0)} & \cellcolor{Gray!10}{\color[HTML]{0072B2}-0.7 (-1.0, 0.0)} & \cellcolor{Gray!10}{\color[HTML]{0072B2}-0.5 (-1.2, 0.2)}\\
66.0 (31.5, 100.5) & -3.4 (-61.7, 54.8) & 4.0 & \color[HTML]{D55E00}48.0 (18.3, 77.6) & \color[HTML]{0072B2}40.2 (-26.0, 79.5) & \color[HTML]{0072B2}52.5 (-19.8, 82.7) & \color[HTML]{0072B2}58.3 (-12.7, 85.7) & \color[HTML]{0072B2}35.4 (-32.2, 103.0)\\
\cellcolor{Gray!10}{-6.8 (-12.9, -0.7)} & \cellcolor{Gray!10}{-0.7 (-1.5, 0.1)} & \cellcolor{Gray!10}{3.8} & \cellcolor{Gray!10}{\color[HTML]{D55E00}-0.8 (-1.6, -0.0)} & \cellcolor{Gray!10}{\color[HTML]{D55E00}-1.4 (-9.1, -0.4)} & \cellcolor{Gray!10}{\color[HTML]{D55E00}-0.8 (-5.9, -0.1)} & \cellcolor{Gray!10}{\color[HTML]{0072B2}-0.7 (-1.6, 0.1)} & \cellcolor{Gray!10}{\color[HTML]{0072B2}-3.0 (-8.8, 2.8)}\\
0.5 (0.2, 0.8) & 0.0 (-0.4, 0.4) & 3.8 & \color[HTML]{D55E00}0.3 (0.1, 0.6) & \color[HTML]{0072B2}0.3 (-0.1, 0.6) & \color[HTML]{0072B2}0.3 (-0.1, 0.7) & \color[HTML]{0072B2}0.4 (-0.1, 0.7) & \color[HTML]{0072B2}0.3 (-0.2, 0.8)\\
\cellcolor{Gray!10}{-0.7 (-1.1, -0.4)} & \cellcolor{Gray!10}{-0.2 (-0.6, 0.3)} & \cellcolor{Gray!10}{3.7} & \cellcolor{Gray!10}{\color[HTML]{D55E00}-0.5 (-0.8, -0.2)} & \cellcolor{Gray!10}{\color[HTML]{0072B2}-0.5 (-0.9, 0.0)} & \cellcolor{Gray!10}{\color[HTML]{D55E00}-0.5 (-0.9, -0.0)} & \cellcolor{Gray!10}{\color[HTML]{D55E00}-0.6 (-0.9, -0.0)} & \cellcolor{Gray!10}{\color[HTML]{0072B2}-0.5 (-1.0, 0.1)}\\
-1.8 (-4.9, 1.4) & 1.4 (0.5, 2.3) & 3.6 & \color[HTML]{D55E00}1.2 (0.4, 2.0) & \color[HTML]{0072B2}0.7 (-3.0, 1.8) & \color[HTML]{0072B2}1.3 (-2.1, 2.0) & \color[HTML]{0072B2}1.4 (-0.6, 2.1) & \color[HTML]{0072B2}0.2 (-2.8, 3.2)\\
\cellcolor{Gray!10}{0.4 (0.1, 0.7)} & \cellcolor{Gray!10}{0.1 (-0.2, 0.3)} & \cellcolor{Gray!10}{3.6} & \cellcolor{Gray!10}{\color[HTML]{D55E00}0.2 (0.0, 0.4)} & \cellcolor{Gray!10}{\color[HTML]{0072B2}0.2 (-0.1, 0.6)} & \cellcolor{Gray!10}{\color[HTML]{0072B2}0.2 (-0.1, 0.6)} & \cellcolor{Gray!10}{\color[HTML]{0072B2}0.2 (-0.1, 0.5)} & \cellcolor{Gray!10}{\color[HTML]{0072B2}0.2 (-0.1, 0.6)}\\
-0.1 (-0.6, 0.5) & -0.8 (-1.4, -0.2) & 3.4 & \color[HTML]{D55E00}-0.4 (-0.8, -0.0) & \color[HTML]{0072B2}-0.4 (-1.0, 0.2) & \color[HTML]{0072B2}-0.4 (-1.0, 0.2) & \color[HTML]{0072B2}-0.4 (-1.0, 0.2) & \color[HTML]{0072B2}-0.4 (-1.2, 0.3)\\
\cellcolor{Gray!10}{1.7 (0.5, 2.8)} & \cellcolor{Gray!10}{0.3 (-0.6, 1.2)} & \cellcolor{Gray!10}{3.3} & \cellcolor{Gray!10}{\color[HTML]{D55E00}1.0 (0.3, 1.8)} & \cellcolor{Gray!10}{\color[HTML]{0072B2}0.9 (-0.0, 2.1)} & \cellcolor{Gray!10}{\color[HTML]{0072B2}0.8 (-0.1, 2.1)} & \cellcolor{Gray!10}{\color[HTML]{0072B2}0.7 (-0.1, 2.0)} & \cellcolor{Gray!10}{\color[HTML]{0072B2}0.9 (-0.4, 2.3)}\\
-4.9 (-9.6, -0.2) & -0.5 (-1.6, 0.6) & 3.2 & \color[HTML]{0072B2}-0.7 (-1.7, 0.3) & \color[HTML]{D55E00}-1.3 (-6.7, -0.0) & \color[HTML]{0072B2}-0.6 (-5.1, 0.2) & \color[HTML]{0072B2}-0.5 (-1.9, 0.4) & \color[HTML]{0072B2}-2.1 (-6.2, 2.1)\\
\cellcolor{Gray!10}{-45.0 (-70.5, -19.5)} & \cellcolor{Gray!10}{-1.8 (-41.7, 38.1)} & \cellcolor{Gray!10}{3.2} & \cellcolor{Gray!10}{\color[HTML]{D55E00}-32.5 (-53.9, -11.0)} & \cellcolor{Gray!10}{\color[HTML]{0072B2}-28.2 (-55.1, 13.7)} & \cellcolor{Gray!10}{\color[HTML]{0072B2}-34.7 (-57.1, 10.1)} & \cellcolor{Gray!10}{\color[HTML]{0072B2}-38.6 (-58.9, 6.0)} & \cellcolor{Gray!10}{\color[HTML]{0072B2}-26.2 (-68.2, 15.7)}\\
-0.0 (-0.2, 0.1) & -0.2 (-0.4, -0.1) & 3.0 & \color[HTML]{D55E00}-0.1 (-0.2, -0.0) & \color[HTML]{0072B2}-0.1 (-0.3, 0.0) & \color[HTML]{0072B2}-0.1 (-0.3, 0.0) & \color[HTML]{0072B2}-0.1 (-0.3, 0.0) & \color[HTML]{0072B2}-0.1 (-0.3, 0.1)\\
\cellcolor{Gray!10}{1.4 (0.4, 2.3)} & \cellcolor{Gray!10}{0.1 (-1.1, 1.2)} & \cellcolor{Gray!10}{3.0} & \cellcolor{Gray!10}{\color[HTML]{D55E00}0.9 (0.2, 1.6)} & \cellcolor{Gray!10}{\color[HTML]{0072B2}0.8 (-0.4, 1.7)} & \cellcolor{Gray!10}{\color[HTML]{0072B2}0.9 (-0.3, 1.8)} & \cellcolor{Gray!10}{\color[HTML]{0072B2}1.0 (-0.3, 1.8)} & \cellcolor{Gray!10}{\color[HTML]{0072B2}0.8 (-0.5, 2.0)}\\
0.4 (0.1, 0.7) & 0.1 (-0.2, 0.3) & 3.0 & \color[HTML]{D55E00}0.2 (0.0, 0.4) & \color[HTML]{0072B2}0.2 (-0.0, 0.5) & \color[HTML]{0072B2}0.2 (-0.0, 0.5) & \color[HTML]{0072B2}0.2 (-0.0, 0.5) & \color[HTML]{0072B2}0.2 (-0.1, 0.5)\\
\cellcolor{Gray!10}{-1.3 (-2.3, -0.3)} & \cellcolor{Gray!10}{-0.3 (-0.9, 0.4)} & \cellcolor{Gray!10}{2.7} & \cellcolor{Gray!10}{\color[HTML]{D55E00}-0.6 (-1.2, -0.1)} & \cellcolor{Gray!10}{\color[HTML]{D55E00}-0.7 (-1.7, -0.0)} & \cellcolor{Gray!10}{\color[HTML]{0072B2}-0.6 (-1.6, 0.0)} & \cellcolor{Gray!10}{\color[HTML]{0072B2}-0.5 (-1.5, 0.1)} & \cellcolor{Gray!10}{\color[HTML]{0072B2}-0.7 (-1.7, 0.2)}\\
0.2 (-0.6, 1.0) & 1.1 (0.5, 1.6) & 2.7 & \color[HTML]{D55E00}0.8 (0.3, 1.3) & \color[HTML]{0072B2}0.7 (-0.1, 1.3) & \color[HTML]{0072B2}0.8 (-0.0, 1.3) & \color[HTML]{D55E00}0.9 (0.0, 1.4) & \color[HTML]{0072B2}0.7 (-0.1, 1.5)\\
\cellcolor{Gray!10}{-1.1 (-1.9, -0.3)} & \cellcolor{Gray!10}{0.0 (-1.0, 1.0)} & \cellcolor{Gray!10}{2.7} & \cellcolor{Gray!10}{\color[HTML]{D55E00}-0.7 (-1.3, -0.0)} & \cellcolor{Gray!10}{\color[HTML]{0072B2}-0.6 (-1.4, 0.4)} & \cellcolor{Gray!10}{\color[HTML]{0072B2}-0.7 (-1.5, 0.4)} & \cellcolor{Gray!10}{\color[HTML]{0072B2}-0.8 (-1.5, 0.3)} & \cellcolor{Gray!10}{\color[HTML]{0072B2}-0.6 (-1.7, 0.5)}\\
0.6 (0.2, 1.0) & 0.1 (-0.3, 0.5) & 2.6 & \color[HTML]{D55E00}0.3 (0.1, 0.6) & \color[HTML]{0072B2}0.3 (-0.1, 0.7) & \color[HTML]{0072B2}0.3 (-0.1, 0.7) & \color[HTML]{0072B2}0.3 (-0.1, 0.7) & \color[HTML]{0072B2}0.3 (-0.1, 0.8)\\
\cellcolor{Gray!10}{-0.6 (-1.0, -0.2)} & \cellcolor{Gray!10}{-0.1 (-0.5, 0.3)} & \cellcolor{Gray!10}{2.6} & \cellcolor{Gray!10}{\color[HTML]{D55E00}-0.3 (-0.6, -0.1)} & \cellcolor{Gray!10}{\color[HTML]{0072B2}-0.3 (-0.7, 0.1)} & \cellcolor{Gray!10}{\color[HTML]{0072B2}-0.3 (-0.7, 0.1)} & \cellcolor{Gray!10}{\color[HTML]{0072B2}-0.3 (-0.7, 0.1)} & \cellcolor{Gray!10}{\color[HTML]{0072B2}-0.3 (-0.8, 0.1)}\\
-0.7 (-1.3, -0.1) & -0.2 (-0.5, 0.2) & 2.3 & \color[HTML]{D55E00}-0.3 (-0.7, -0.0) & \color[HTML]{0072B2}-0.4 (-0.9, 0.0) & \color[HTML]{0072B2}-0.3 (-0.9, 0.0) & \color[HTML]{0072B2}-0.3 (-0.8, 0.1) & \color[HTML]{0072B2}-0.4 (-0.9, \vphantom{1} 0.1)\\
\cellcolor{Gray!10}{-0.7 (-1.3, -0.1)} & \cellcolor{Gray!10}{-0.2 (-0.5, 0.2)} & \cellcolor{Gray!10}{2.3} & \cellcolor{Gray!10}{\color[HTML]{D55E00}-0.3 (-0.7, -0.0)} & \cellcolor{Gray!10}{\color[HTML]{0072B2}-0.4 (-0.9, 0.0)} & \cellcolor{Gray!10}{\color[HTML]{0072B2}-0.3 (-0.9, 0.0)} & \cellcolor{Gray!10}{\color[HTML]{0072B2}-0.3 (-0.8, 0.1)} & \cellcolor{Gray!10}{\color[HTML]{0072B2}-0.4 (-0.9, 0.1)}\\
0.8 (0.1, 1.4) & 0.0 (-0.7, 0.8) & 2.2 & \color[HTML]{D55E00}0.5 (0.0, 1.0) & \color[HTML]{0072B2}0.4 (-0.2, 1.0) & \color[HTML]{0072B2}0.5 (-0.2, 1.1) & \color[HTML]{0072B2}0.5 (-0.2, 1.1) & \color[HTML]{0072B2}0.4 (-0.3, 1.2)\\
\cellcolor{Gray!10}{-0.2 (-0.9, 0.5)} & \cellcolor{Gray!10}{-0.9 (-1.6, -0.2)} & \cellcolor{Gray!10}{2.2} & \cellcolor{Gray!10}{\color[HTML]{D55E00}-0.5 (-1.0, -0.1)} & \cellcolor{Gray!10}{\color[HTML]{0072B2}-0.5 (-1.2, 0.1)} & \cellcolor{Gray!10}{\color[HTML]{0072B2}-0.5 (-1.2, 0.1)} & \cellcolor{Gray!10}{\color[HTML]{0072B2}-0.5 (-1.2, 0.1)} & \cellcolor{Gray!10}{\color[HTML]{0072B2}-0.5 (-1.3, 0.2)}\\
-1.8 (-4.9, 1.4) & 0.6 (0.1, 1.1) & 2.1 & \color[HTML]{D55E00}0.5 (0.1, 1.0) & \color[HTML]{0072B2}0.3 (-3.0, 0.8) & \color[HTML]{0072B2}0.6 (-1.4, 0.9) & \color[HTML]{D55E00}0.6 (0.1, 1.0) & \color[HTML]{0072B2}-0.1 (-2.1, 2.0)\\
\cellcolor{Gray!10}{0.7 (0.3, 1.0)} & \cellcolor{Gray!10}{-0.1 (-1.1, 0.9)} & \cellcolor{Gray!10}{2.1} & \cellcolor{Gray!10}{\color[HTML]{D55E00}0.6 (0.2, 0.9)} & \cellcolor{Gray!10}{\color[HTML]{0072B2}0.5 (-0.5, 0.8)} & \cellcolor{Gray!10}{\color[HTML]{0072B2}0.6 (-0.3, 0.9)} & \cellcolor{Gray!10}{\color[HTML]{D55E00}0.6 (0.0, 0.9)} & \cellcolor{Gray!10}{\color[HTML]{0072B2}0.4 (-0.3, 1.1)}\\
-2.5 (-5.6, 0.6) & -0.2 (-0.6, 0.2) & 2.1 & \color[HTML]{0072B2}-0.2 (-0.6, 0.1) & \color[HTML]{D55E00}-0.4 (-3.7, -0.0) & \color[HTML]{0072B2}-0.2 (-1.9, 0.1) & \color[HTML]{0072B2}-0.2 (-0.6, 0.2) & \color[HTML]{0072B2}-0.8 (-2.8, 1.2)\\
\cellcolor{Gray!10}{-0.4 (-1.5, 0.7)} & \cellcolor{Gray!10}{-2.1 (-4.1, -0.1)} & \cellcolor{Gray!10}{2.0} & \cellcolor{Gray!10}{\color[HTML]{D55E00}-1.0 (-1.9, -0.1)} & \cellcolor{Gray!10}{\color[HTML]{0072B2}-1.0 (-2.9, 0.1)} & \cellcolor{Gray!10}{\color[HTML]{0072B2}-0.7 (-2.6, 0.2)} & \cellcolor{Gray!10}{\color[HTML]{0072B2}-0.6 (-2.3, 0.3)} & \cellcolor{Gray!10}{\color[HTML]{0072B2}-1.0 (-2.6, 0.6)}\\
2.6 (-8.0, 13.2) & 12.6 (3.2, 21.9) & 1.9 & \color[HTML]{D55E00}8.2 (1.2, 15.2) & \color[HTML]{0072B2}7.9 (-1.5, 16.3) & \color[HTML]{0072B2}8.3 (-1.3, 16.5) & \color[HTML]{0072B2}8.7 (-1.0, 16.7) & \color[HTML]{0072B2}7.9 (-1.8, 17.7)\\
\cellcolor{Gray!10}{0.9 (0.2, 1.5)} & \cellcolor{Gray!10}{0.2 (-0.3, 0.8)} & \cellcolor{Gray!10}{1.9} & \cellcolor{Gray!10}{\color[HTML]{D55E00}0.5 (0.1, 0.9)} & \cellcolor{Gray!10}{\color[HTML]{D55E00}0.5 (0.0, 1.1)} & \cellcolor{Gray!10}{\color[HTML]{D55E00}0.5 (0.0, 1.1)} & \cellcolor{Gray!10}{\color[HTML]{0072B2}0.4 (-0.0, 1.1)} & \cellcolor{Gray!10}{\color[HTML]{0072B2}0.5 (-0.1, 1.1)}\\
-1.2 (-2.2, -0.3) & -0.3 (-1.2, 0.6) & 1.9 & \color[HTML]{D55E00}-0.8 (-1.4, -0.1) & \color[HTML]{0072B2}-0.8 (-1.6, 0.0) & \color[HTML]{0072B2}-0.7 (-1.6, 0.1) & \color[HTML]{0072B2}-0.7 (-1.6, 0.1) & \color[HTML]{0072B2}-0.8 (-1.7, 0.1)\\
\cellcolor{Gray!10}{-0.5 (-1.4, 0.5)} & \cellcolor{Gray!10}{0.2 (0.1, 0.3)} & \cellcolor{Gray!10}{1.8} & \cellcolor{Gray!10}{\color[HTML]{D55E00}0.2 (0.0, 0.3)} & \cellcolor{Gray!10}{\color[HTML]{0072B2}0.1 (-0.8, 0.3)} & \cellcolor{Gray!10}{\color[HTML]{0072B2}0.2 (-0.3, 0.3)} & \cellcolor{Gray!10}{\color[HTML]{D55E00}0.2 (0.0, 0.3)} & \cellcolor{Gray!10}{\color[HTML]{0072B2}0.0 (-0.5, 0.6)}\\
-1.9 (-3.5, -0.2) & -0.5 (-1.6, 0.5) & 1.8 & \color[HTML]{D55E00}-0.9 (-1.8, -0.1) & \color[HTML]{D55E00}-1.1 (-2.5, -0.1) & \color[HTML]{D55E00}-0.9 (-2.4, -0.0) & \color[HTML]{0072B2}-0.8 (-2.2, 0.0) & \color[HTML]{0072B2}-1.1 (-2.3, 0.2)\\
\cellcolor{Gray!10}{-0.1 (-0.5, 0.3)} & \cellcolor{Gray!10}{-0.5 (-0.9, -0.1)} & \cellcolor{Gray!10}{1.5} & \cellcolor{Gray!10}{\color[HTML]{D55E00}-0.3 (-0.6, -0.0)} & \cellcolor{Gray!10}{\color[HTML]{0072B2}-0.3 (-0.7, 0.0)} & \cellcolor{Gray!10}{\color[HTML]{0072B2}-0.3 (-0.7, 0.0)} & \cellcolor{Gray!10}{\color[HTML]{0072B2}-0.3 (-0.7, 0.0)} & \cellcolor{Gray!10}{\color[HTML]{0072B2}-0.3 (-0.7, 0.0)}\\
0.2 (-0.6, 1.1) & 1.0 (0.1, 1.8) & 1.4 & \color[HTML]{D55E00}0.6 (0.1, 1.2) & \color[HTML]{0072B2}0.6 (-0.1, 1.3) & \color[HTML]{0072B2}0.6 (-0.1, 1.3) & \color[HTML]{0072B2}0.6 (-0.1, 1.3) & \color[HTML]{0072B2}0.6 (-0.1, 1.3)\\
\cellcolor{Gray!10}{1.2 (-2.7, 5.1)} & \cellcolor{Gray!10}{-1.2 (-2.1, -0.3)} & \cellcolor{Gray!10}{1.4} & \cellcolor{Gray!10}{\color[HTML]{D55E00}-1.1 (-1.9, -0.2)} & \cellcolor{Gray!10}{\color[HTML]{0072B2}-0.8 (-1.7, 2.7)} & \cellcolor{Gray!10}{\color[HTML]{0072B2}-1.1 (-1.8, 1.3)} & \cellcolor{Gray!10}{\color[HTML]{D55E00}-1.2 (-1.9, -0.2)} & \cellcolor{Gray!10}{\color[HTML]{0072B2}-0.8 (-2.6, 1.0)}\\
-0.0 (-0.2, 0.1) & -0.2 (-0.3, -0.0) & 1.3 & \color[HTML]{D55E00}-0.1 (-0.2, -0.0) & \color[HTML]{0072B2}-0.1 (-0.2, 0.0) & \color[HTML]{0072B2}-0.1 (-0.2, 0.0) & \color[HTML]{0072B2}-0.1 (-0.2, 0.0) & \color[HTML]{0072B2}-0.1 (-0.2, 0.0)\\
\cellcolor{Gray!10}{0.1 (-0.2, 0.4)} & \cellcolor{Gray!10}{0.3 (0.1, 0.6)} & \cellcolor{Gray!10}{1.3} & \cellcolor{Gray!10}{\color[HTML]{D55E00}0.2 (0.0, 0.4)} & \cellcolor{Gray!10}{\color[HTML]{0072B2}0.2 (-0.0, 0.4)} & \cellcolor{Gray!10}{\color[HTML]{0072B2}0.2 (-0.0, 0.4)} & \cellcolor{Gray!10}{\color[HTML]{0072B2}0.2 (-0.0, 0.4)} & \cellcolor{Gray!10}{\color[HTML]{D55E00}0.2 (0.0, 0.5)}\\
0.1 (-0.5, 0.7) & 0.6 (0.1, 1.0) & 1.3 & \color[HTML]{D55E00}0.4 (0.0, 0.7) & \color[HTML]{0072B2}0.4 (-0.1, 0.8) & \color[HTML]{0072B2}0.4 (-0.1, 0.8) & \color[HTML]{0072B2}0.4 (-0.1, 0.8) & \color[HTML]{0072B2}0.4 (-0.1, 0.8)\\
\cellcolor{Gray!10}{1.1 (-2.1, 4.4)} & \cellcolor{Gray!10}{3.2 (1.8, 4.6)} & \cellcolor{Gray!10}{1.3} & \cellcolor{Gray!10}{\color[HTML]{D55E00}3.0 (1.7, 4.3)} & \cellcolor{Gray!10}{\color[HTML]{0072B2}2.6 (-0.1, 3.9)} & \cellcolor{Gray!10}{\color[HTML]{D55E00}2.9 (0.4, 4.0)} & \cellcolor{Gray!10}{\color[HTML]{D55E00}3.1 (1.1, 4.2)} & \cellcolor{Gray!10}{\color[HTML]{D55E00}2.7 (1.0, 4.4)}\\
0.4 (-0.9, 1.7) & 1.5 (0.0, 3.0) & 1.2 & \color[HTML]{D55E00}1.0 (0.1, 1.9) & \color[HTML]{0072B2}0.9 (-0.1, 2.1) & \color[HTML]{0072B2}0.9 (-0.2, 2.1) & \color[HTML]{0072B2}0.8 (-0.2, 2.0) & \color[HTML]{0072B2}0.9 (-0.2, 2.0)\\
\cellcolor{Gray!10}{0.9 (-1.6, 3.3)} & \cellcolor{Gray!10}{2.8 (0.4, 5.2)} & \cellcolor{Gray!10}{1.2} & \cellcolor{Gray!10}{\color[HTML]{D55E00}1.9 (0.1, 3.6)} & \cellcolor{Gray!10}{\color[HTML]{0072B2}1.8 (-0.1, 3.8)} & \cellcolor{Gray!10}{\color[HTML]{0072B2}1.9 (-0.1, 3.8)} & \cellcolor{Gray!10}{\color[HTML]{0072B2}1.9 (-0.1, 3.8)} & \cellcolor{Gray!10}{\color[HTML]{0072B2}1.9 (-0.0, 3.7)}\\
-0.3 (-0.5, -0.0) & 0.0 (-0.5, 0.5) & 1.2 & \color[HTML]{D55E00}-0.2 (-0.4, -0.0) & \color[HTML]{0072B2}-0.2 (-0.4, 0.2) & \color[HTML]{0072B2}-0.2 (-0.4, 0.2) & \color[HTML]{0072B2}-0.2 (-0.4, 0.1) & \color[HTML]{0072B2}-0.2 (-0.5, 0.1)\\
\cellcolor{Gray!10}{2.1 (-1.0, 5.2)} & \cellcolor{Gray!10}{0.3 (-0.2, 0.8)} & \cellcolor{Gray!10}{1.2} & \cellcolor{Gray!10}{\color[HTML]{0072B2}0.4 (-0.1, 0.8)} & \cellcolor{Gray!10}{\color[HTML]{D55E00}0.5 (0.0, 3.3)} & \cellcolor{Gray!10}{\color[HTML]{0072B2}0.3 (-0.1, 1.8)} & \cellcolor{Gray!10}{\color[HTML]{0072B2}0.3 (-0.1, 0.8)} & \cellcolor{Gray!10}{\color[HTML]{0072B2}0.5 (-0.6, 1.5)}\\
0.5 (0.1, 0.9) & 0.2 (-0.4, 0.7) & 1.0 & \color[HTML]{D55E00}0.4 (0.1, 0.7) & \color[HTML]{0072B2}0.4 (-0.1, 0.7) & \color[HTML]{0072B2}0.4 (-0.0, 0.7) & \color[HTML]{D55E00}0.4 (0.0, 0.7) & \color[HTML]{D55E00}0.4 (0.1, 0.7)\\
\cellcolor{Gray!10}{-0.6 (-3.1, 1.9)} & \cellcolor{Gray!10}{-2.0 (-3.1, -0.9)} & \cellcolor{Gray!10}{1.0} & \cellcolor{Gray!10}{\color[HTML]{D55E00}-1.8 (-2.8, -0.8)} & \cellcolor{Gray!10}{\color[HTML]{0072B2}-1.6 (-2.6, 0.4)} & \cellcolor{Gray!10}{\color[HTML]{D55E00}-1.8 (-2.7, -0.1)} & \cellcolor{Gray!10}{\color[HTML]{D55E00}-1.9 (-2.8, -0.5)} & \cellcolor{Gray!10}{\color[HTML]{D55E00}-1.8 (-2.8, -0.8)}\\
0.1 (-0.2, 0.4) & 0.3 (0.0, 0.5) & 0.9 & \color[HTML]{D55E00}0.2 (0.0, 0.4) & \color[HTML]{0072B2}0.2 (-0.0, 0.4) & \color[HTML]{0072B2}0.2 (-0.0, 0.4) & \color[HTML]{0072B2}0.2 (-0.0, 0.4) & \color[HTML]{D55E00}0.2 (0.0, 0.4)\\
\cellcolor{Gray!10}{0.6 (-0.1, 1.3)} & \cellcolor{Gray!10}{0.2 (-0.2, 0.6)} & \cellcolor{Gray!10}{0.9} & \cellcolor{Gray!10}{\color[HTML]{0072B2}0.3 (-0.0, 0.6)} & \cellcolor{Gray!10}{\color[HTML]{D55E00}0.3 (0.0, 0.9)} & \cellcolor{Gray!10}{\color[HTML]{0072B2}0.3 (-0.0, 0.8)} & \cellcolor{Gray!10}{\color[HTML]{0072B2}0.3 (-0.0, 0.7)} & \cellcolor{Gray!10}{\color[HTML]{0072B2}0.3 (-0.0, 0.6)}\\
-0.1 (-0.6, 0.3) & -0.4 (-0.6, -0.1) & 0.8 & \color[HTML]{D55E00}-0.3 (-0.5, -0.1) & \color[HTML]{0072B2}-0.3 (-0.5, 0.0) & \color[HTML]{0072B2}-0.3 (-0.5, 0.0) & \color[HTML]{D55E00}-0.3 (-0.5, -0.0) & \color[HTML]{D55E00}-0.3 (-0.5, -0.1)\\
\cellcolor{Gray!10}{0.1 (0.0, 0.3)} & \cellcolor{Gray!10}{0.0 (-0.1, 0.2)} & \cellcolor{Gray!10}{0.8} & \cellcolor{Gray!10}{\color[HTML]{D55E00}0.1 (0.0, 0.2)} & \cellcolor{Gray!10}{\color[HTML]{0072B2}0.1 (-0.0, 0.2)} & \cellcolor{Gray!10}{\color[HTML]{0072B2}0.1 (-0.0, 0.2)} & \cellcolor{Gray!10}{\color[HTML]{0072B2}0.1 (-0.0, 0.2)} & \cellcolor{Gray!10}{\color[HTML]{0072B2}0.1 (-0.0, 0.2)}\\
-5.2 (-9.7, -0.7) & -1.9 (-8.2, 4.4) & 0.7 & \color[HTML]{D55E00}-4.1 (-7.7, -0.5) & \color[HTML]{0072B2}-3.8 (-7.3, 0.6) & \color[HTML]{0072B2}-4.1 (-7.5, 0.2) & \color[HTML]{D55E00}-4.4 (-7.7, -0.1) & \color[HTML]{D55E00}-4.1 (-7.7, -0.5)\\
\cellcolor{Gray!10}{-0.3 (-0.5, -0.1)} & \cellcolor{Gray!10}{-0.1 (-0.5, 0.2)} & \cellcolor{Gray!10}{0.7} & \cellcolor{Gray!10}{\color[HTML]{D55E00}-0.2 (-0.4, -0.1)} & \cellcolor{Gray!10}{\color[HTML]{0072B2}-0.2 (-0.4, 0.0)} & \cellcolor{Gray!10}{\color[HTML]{D55E00}-0.2 (-0.4, -0.0)} & \cellcolor{Gray!10}{\color[HTML]{D55E00}-0.3 (-0.4, -0.0)} & \cellcolor{Gray!10}{\color[HTML]{D55E00}-0.2 (-0.4, -0.1)}\\
0.4 (-0.5, 1.2) & 0.9 (0.1, 1.7) & 0.7 & \color[HTML]{D55E00}0.6 (0.0, 1.2) & \color[HTML]{0072B2}0.6 (-0.0, 1.3) & \color[HTML]{D55E00}0.6 (0.0, 1.3) & \color[HTML]{D55E00}0.6 (0.0, 1.3) & \color[HTML]{D55E00}0.6 (0.0, 1.2)\\
\cellcolor{Gray!10}{-1.5 (-3.8, 0.9)} & \cellcolor{Gray!10}{-0.4 (-1.1, 0.2)} & \cellcolor{Gray!10}{0.7} & \cellcolor{Gray!10}{\color[HTML]{0072B2}-0.5 (-1.1, 0.1)} & \cellcolor{Gray!10}{\color[HTML]{D55E00}-0.7 (-2.4, -0.0)} & \cellcolor{Gray!10}{\color[HTML]{D55E00}-0.5 (-1.7, -0.0)} & \cellcolor{Gray!10}{\color[HTML]{0072B2}-0.5 (-1.1, 0.1)} & \cellcolor{Gray!10}{\color[HTML]{0072B2}-0.5 (-1.1, 0.1)}\\
-0.6 (-1.4, 0.1) & -0.2 (-0.7, 0.2) & 0.7 & \color[HTML]{0072B2}-0.4 (-0.8, 0.0) & \color[HTML]{D55E00}-0.4 (-0.9, -0.0) & \color[HTML]{0072B2}-0.3 (-0.9, 0.0) & \color[HTML]{0072B2}-0.3 (-0.8, 0.0) & \color[HTML]{0072B2}-0.4 (-0.8, 0.0)\\
\cellcolor{Gray!10}{-0.5 (-0.9, -0.1)} & \cellcolor{Gray!10}{-0.2 (-0.8, 0.3)} & \cellcolor{Gray!10}{0.6} & \cellcolor{Gray!10}{\color[HTML]{D55E00}-0.4 (-0.7, -0.1)} & \cellcolor{Gray!10}{\color[HTML]{0072B2}-0.4 (-0.7, 0.0)} & \cellcolor{Gray!10}{\color[HTML]{D55E00}-0.4 (-0.7, -0.0)} & \cellcolor{Gray!10}{\color[HTML]{D55E00}-0.4 (-0.7, -0.0)} & \cellcolor{Gray!10}{\color[HTML]{D55E00}-0.4 (-0.7, -0.1)}\\
0.0 (-0.4, 0.4) & -0.1 (-0.2, -0.0) & 0.6 & \color[HTML]{D55E00}-0.1 (-0.2, -0.0) & \color[HTML]{0072B2}-0.1 (-0.2, 0.2) & \color[HTML]{0072B2}-0.1 (-0.2, 0.0) & \color[HTML]{D55E00}-0.1 (-0.2, -0.0) & \color[HTML]{D55E00}-0.1 (-0.2, -0.0)\\
\cellcolor{Gray!10}{0.0 (-0.7, 0.8)} & \cellcolor{Gray!10}{0.3 (0.1, 0.6)} & \cellcolor{Gray!10}{0.6} & \cellcolor{Gray!10}{\color[HTML]{D55E00}0.3 (0.1, 0.6)} & \cellcolor{Gray!10}{\color[HTML]{0072B2}0.3 (-0.3, 0.5)} & \cellcolor{Gray!10}{\color[HTML]{0072B2}0.3 (-0.1, 0.5)} & \cellcolor{Gray!10}{\color[HTML]{D55E00}0.3 (0.1, 0.5)} & \cellcolor{Gray!10}{\color[HTML]{D55E00}0.3 (0.1, 0.6)}\\
-0.3 (-0.5, -0.0) & -0.1 (-0.5, 0.4) & 0.6 & \color[HTML]{D55E00}-0.2 (-0.4, -0.0) & \color[HTML]{0072B2}-0.2 (-0.4, 0.1) & \color[HTML]{0072B2}-0.2 (-0.4, 0.0) & \color[HTML]{0072B2}-0.2 (-0.4, 0.0) & \color[HTML]{D55E00}-0.2 (-0.4, -0.0)\\
\cellcolor{Gray!10}{-0.4 (-2.8, 2.0)} & \cellcolor{Gray!10}{-1.4 (-2.2, -0.6)} & \cellcolor{Gray!10}{0.6} & \cellcolor{Gray!10}{\color[HTML]{D55E00}-1.3 (-2.1, -0.5)} & \cellcolor{Gray!10}{\color[HTML]{0072B2}-1.2 (-1.9, 0.5)} & \cellcolor{Gray!10}{\color[HTML]{D55E00}-1.3 (-1.9, -0.0)} & \cellcolor{Gray!10}{\color[HTML]{D55E00}-1.4 (-2.0, -0.5)} & \cellcolor{Gray!10}{\color[HTML]{D55E00}-1.3 (-2.1, -0.5)}\\
-1.0 (-1.9, -0.1) & -0.4 (-1.7, 0.8) & 0.6 & \color[HTML]{D55E00}-0.8 (-1.5, -0.1) & \color[HTML]{0072B2}-0.8 (-1.5, 0.1) & \color[HTML]{0072B2}-0.8 (-1.5, 0.0) & \color[HTML]{D55E00}-0.9 (-1.5, -0.0) & \color[HTML]{D55E00}-0.8 (-1.5, -0.1)\\
\cellcolor{Gray!10}{-0.7 (-1.6, 0.2)} & \cellcolor{Gray!10}{-0.3 (-0.8, 0.2)} & \cellcolor{Gray!10}{0.5} & \cellcolor{Gray!10}{\color[HTML]{0072B2}-0.4 (-0.8, 0.0)} & \cellcolor{Gray!10}{\color[HTML]{D55E00}-0.4 (-1.0, -0.0)} & \cellcolor{Gray!10}{\color[HTML]{D55E00}-0.4 (-0.9, -0.0)} & \cellcolor{Gray!10}{\color[HTML]{0072B2}-0.3 (-0.8, 0.0)} & \cellcolor{Gray!10}{\color[HTML]{0072B2}-0.4 (-0.8, 0.0)}\\
-0.1 (-0.3, 0.1) & -0.3 (-0.7, 0.1) & 0.5 & \color[HTML]{0072B2}-0.2 (-0.4, 0.0) & \color[HTML]{D55E00}-0.2 (-0.5, -0.0) & \color[HTML]{0072B2}-0.2 (-0.4, 0.0) & \color[HTML]{0072B2}-0.1 (-0.4, 0.0) & \color[HTML]{0072B2}-0.2 (-0.4, 0.0)\\
\cellcolor{Gray!10}{-2.4 (-3.2, -1.5)} & \cellcolor{Gray!10}{-1.1 (-4.4, 2.1)} & \cellcolor{Gray!10}{0.5} & \cellcolor{Gray!10}{\color[HTML]{D55E00}-2.3 (-3.1, -1.4)} & \cellcolor{Gray!10}{\color[HTML]{0072B2}-2.1 (-3.0, 0.1)} & \cellcolor{Gray!10}{\color[HTML]{D55E00}-2.3 (-3.0, -0.8)} & \cellcolor{Gray!10}{\color[HTML]{D55E00}-2.3 (-3.1, -1.4)} & \cellcolor{Gray!10}{\color[HTML]{D55E00}-2.3 (-3.1, -1.4)}\\
-1.6 (-3.4, 0.2) & -0.8 (-2.5, 0.9) & 0.4 & \color[HTML]{0072B2}-1.2 (-2.4, 0.1) & \color[HTML]{D55E00}-1.2 (-2.4, -0.0) & \color[HTML]{D55E00}-1.2 (-2.4, -0.0) & \color[HTML]{0072B2}-1.1 (-2.4, 0.0) & \color[HTML]{0072B2}-1.2 (-2.4, 0.1)\\
\cellcolor{Gray!10}{1.6 (-0.2, 3.4)} & \cellcolor{Gray!10}{0.8 (-0.9, 2.5)} & \cellcolor{Gray!10}{0.4} & \cellcolor{Gray!10}{\color[HTML]{0072B2}1.2 (-0.1, 2.4)} & \cellcolor{Gray!10}{\color[HTML]{D55E00}1.2 (0.0, 2.4)} & \cellcolor{Gray!10}{\color[HTML]{D55E00}1.2 (0.0, 2.4)} & \cellcolor{Gray!10}{\color[HTML]{0072B2}1.1 (-0.0, 2.4)} & \cellcolor{Gray!10}{\color[HTML]{0072B2}1.2 (-0.1, 2.4)}\\
17.0 (-64.3, 98.3) & 44.0 (14.4, 73.5) & 0.4 & \color[HTML]{D55E00}40.8 (13.0, 68.6) & \color[HTML]{0072B2}36.8 (-14.5, 63.8) & \color[HTML]{D55E00}40.9 (2.4, 64.3) & \color[HTML]{D55E00}42.8 (13.0, 66.7) & \color[HTML]{D55E00}40.8 (13.0, 68.6)\\
\cellcolor{Gray!10}{0.6 (-0.0, 1.2)} & \cellcolor{Gray!10}{0.3 (-0.5, 1.1)} & \cellcolor{Gray!10}{0.4} & \cellcolor{Gray!10}{\color[HTML]{D55E00}0.5 (0.0, 1.0)} & \cellcolor{Gray!10}{\color[HTML]{0072B2}0.5 (-0.0, 0.9)} & \cellcolor{Gray!10}{\color[HTML]{0072B2}0.5 (-0.0, 0.9)} & \cellcolor{Gray!10}{\color[HTML]{D55E00}0.5 (0.0, 1.0)} & \cellcolor{Gray!10}{\color[HTML]{D55E00}0.5 (0.0, 1.0)}\\
0.4 (-0.3, 1.2) & 0.8 (-0.1, 1.7) & 0.4 & \color[HTML]{0072B2}0.6 (-0.0, 1.2) & \color[HTML]{D55E00}0.6 (0.1, 1.2) & \color[HTML]{D55E00}0.6 (0.0, 1.2) & \color[HTML]{D55E00}0.6 (0.0, 1.1) & \color[HTML]{0072B2}0.6 (-0.0, 1.2)\\
\cellcolor{Gray!10}{-0.2 (-0.3, -0.0)} & \cellcolor{Gray!10}{0.1 (-0.7, 0.8)} & \cellcolor{Gray!10}{0.4} & \cellcolor{Gray!10}{\color[HTML]{D55E00}-0.2 (-0.3, -0.0)} & \cellcolor{Gray!10}{\color[HTML]{0072B2}-0.1 (-0.3, 0.3)} & \cellcolor{Gray!10}{\color[HTML]{0072B2}-0.2 (-0.3, 0.1)} & \cellcolor{Gray!10}{\color[HTML]{D55E00}-0.2 (-0.3, -0.0)} & \cellcolor{Gray!10}{\color[HTML]{D55E00}-0.2 (-0.3, -0.0)}\\
-0.2 (-0.6, 0.2) & -0.3 (-0.8, 0.1) & 0.3 & \color[HTML]{0072B2}-0.3 (-0.5, 0.0) & \color[HTML]{D55E00}-0.3 (-0.5, -0.0) & \color[HTML]{D55E00}-0.3 (-0.5, -0.0) & \color[HTML]{D55E00}-0.3 (-0.5, -0.0) & \color[HTML]{0072B2}-0.3 (-0.5, 0.0)\\
\cellcolor{Gray!10}{0.3 (-0.4, 1.0)} & \cellcolor{Gray!10}{0.5 (-0.0, 1.0)} & \cellcolor{Gray!10}{0.3} & \cellcolor{Gray!10}{\color[HTML]{0072B2}0.4 (-0.0, 0.8)} & \cellcolor{Gray!10}{\color[HTML]{0072B2}0.4 (-0.0, 0.8)} & \cellcolor{Gray!10}{\color[HTML]{0072B2}0.4 (-0.0, 0.8)} & \cellcolor{Gray!10}{\color[HTML]{D55E00}0.4 (0.0, 0.8)} & \cellcolor{Gray!10}{\color[HTML]{0072B2}0.4 (-0.0, 0.8)}\\
-2.3 (-4.4, -0.1) & -1.1 (-4.3, 2.1) & 0.3 & \color[HTML]{D55E00}-1.9 (-3.7, -0.2) & \color[HTML]{0072B2}-1.8 (-3.4, 0.2) & \color[HTML]{0072B2}-1.9 (-3.5, 0.0) & \color[HTML]{D55E00}-2.0 (-3.6, -0.1) & \color[HTML]{D55E00}-1.9 (-3.7, -0.1)\\
\cellcolor{Gray!10}{-1.0 (-4.9, 2.9)} & \cellcolor{Gray!10}{-2.2 (-4.2, -0.3)} & \cellcolor{Gray!10}{0.3} & \cellcolor{Gray!10}{\color[HTML]{D55E00}-2.0 (-3.7, -0.2)} & \cellcolor{Gray!10}{\color[HTML]{0072B2}-1.8 (-3.4, 0.5)} & \cellcolor{Gray!10}{\color[HTML]{0072B2}-2.0 (-3.5, 0.1)} & \cellcolor{Gray!10}{\color[HTML]{D55E00}-2.1 (-3.6, -0.2)} & \cellcolor{Gray!10}{\color[HTML]{D55E00}-2.0 (-3.7, -0.2)}\\
-0.4 (-0.9, 0.1) & -0.2 (-0.7, 0.2) & 0.2 & \color[HTML]{0072B2}-0.3 (-0.6, 0.0) & \color[HTML]{D55E00}-0.3 (-0.6, -0.0) & \color[HTML]{D55E00}-0.3 (-0.6, -0.0) & \color[HTML]{D55E00}-0.3 (-0.6, -0.0) & \color[HTML]{0072B2}-0.3 (-0.6, 0.0)\\
\cellcolor{Gray!10}{0.1 (-0.2, 0.4)} & \cellcolor{Gray!10}{0.2 (-0.0, 0.5)} & \cellcolor{Gray!10}{0.2} & \cellcolor{Gray!10}{\color[HTML]{0072B2}0.2 (-0.0, 0.4)} & \cellcolor{Gray!10}{\color[HTML]{0072B2}0.2 (-0.0, 0.4)} & \cellcolor{Gray!10}{\color[HTML]{0072B2}0.2 (-0.0, 0.4)} & \cellcolor{Gray!10}{\color[HTML]{D55E00}0.2 (0.0, 0.4)} & \cellcolor{Gray!10}{\color[HTML]{0072B2}0.2 (-0.0, 0.4)}\\
0.2 (-0.1, 0.5) & 0.1 (-0.1, 0.4) & 0.2 & \color[HTML]{0072B2}0.2 (-0.0, 0.3) & \color[HTML]{D55E00}0.2 (0.0, 0.4) & \color[HTML]{D55E00}0.2 (0.0, 0.3) & \color[HTML]{0072B2}0.2 (-0.0, 0.3) & \color[HTML]{0072B2}0.2 (-0.0, 0.3)\\
\cellcolor{Gray!10}{-0.2 (-0.7, 0.2)} & \cellcolor{Gray!10}{-0.1 (-0.3, 0.0)} & \cellcolor{Gray!10}{0.2} & \cellcolor{Gray!10}{\color[HTML]{0072B2}-0.1 (-0.3, 0.0)} & \cellcolor{Gray!10}{\color[HTML]{D55E00}-0.2 (-0.4, -0.0)} & \cellcolor{Gray!10}{\color[HTML]{D55E00}-0.1 (-0.3, -0.0)} & \cellcolor{Gray!10}{\color[HTML]{0072B2}-0.1 (-0.3, 0.0)} & \cellcolor{Gray!10}{\color[HTML]{0072B2}-0.1 (-0.3, 0.0)}\\
5.0 (-2.8, 12.8) & 8.2 (-2.7, 19.1) & 0.2 & \color[HTML]{0072B2}6.1 (-0.3, 12.4) & \color[HTML]{D55E00}6.3 (0.6, 12.9) & \color[HTML]{D55E00}6.1 (0.5, 12.5) & \color[HTML]{D55E00}5.8 (0.3, 12.1) & \color[HTML]{0072B2}6.1 (-0.3, 12.4)\\
\cellcolor{Gray!10}{0.4 (0.0, 0.8)} & \cellcolor{Gray!10}{0.1 (-1.4, 1.7)} & \cellcolor{Gray!10}{0.1} & \cellcolor{Gray!10}{\color[HTML]{D55E00}0.4 (0.0, 0.8)} & \cellcolor{Gray!10}{\color[HTML]{0072B2}0.4 (-0.5, 0.8)} & \cellcolor{Gray!10}{\color[HTML]{0072B2}0.4 (-0.0, 0.7)} & \cellcolor{Gray!10}{\color[HTML]{D55E00}0.4 (0.0, 0.8)} & \cellcolor{Gray!10}{\color[HTML]{D55E00}0.4 (0.0, 0.8)}\\
-0.1 (-0.5, 0.3) & -0.1 (-0.2, -0.0) & 0.1 & \color[HTML]{D55E00}-0.1 (-0.2, -0.0) & \color[HTML]{0072B2}-0.1 (-0.2, 0.1) & \color[HTML]{D55E00}-0.1 (-0.2, -0.0) & \color[HTML]{D55E00}-0.1 (-0.2, -0.0) & \color[HTML]{D55E00}-0.1 (-0.2, -0.0)\\
\cellcolor{Gray!10}{0.6 (-0.2, 1.3)} & \cellcolor{Gray!10}{0.4 (-0.3, 1.1)} & \cellcolor{Gray!10}{0.1} & \cellcolor{Gray!10}{\color[HTML]{0072B2}0.5 (-0.0, 1.0)} & \cellcolor{Gray!10}{\color[HTML]{D55E00}0.5 (0.0, 1.0)} & \cellcolor{Gray!10}{\color[HTML]{D55E00}0.5 (0.0, 1.0)} & \cellcolor{Gray!10}{\color[HTML]{D55E00}0.5 (0.0, 1.0)} & \cellcolor{Gray!10}{\color[HTML]{0072B2}0.5 (-0.0, 1.0)}\\
-1.1 (-4.3, 2.1) & -0.7 (-1.5, 0.0) & 0.1 & \color[HTML]{D55E00}-0.8 (-1.5, -0.0) & \color[HTML]{0072B2}-0.8 (-2.4, 0.2) & \color[HTML]{D55E00}-0.8 (-1.6, -0.1) & \color[HTML]{D55E00}-0.7 (-1.5, -0.0) & \color[HTML]{D55E00}-0.8 (-1.5, -0.0)\\
\cellcolor{Gray!10}{0.6 (0.1, 1.1)} & \cellcolor{Gray!10}{0.4 (-0.6, 1.5)} & \cellcolor{Gray!10}{0.1} & \cellcolor{Gray!10}{\color[HTML]{D55E00}0.5 (0.1, 1.0)} & \cellcolor{Gray!10}{\color[HTML]{0072B2}0.5 (-0.0, 1.0)} & \cellcolor{Gray!10}{\color[HTML]{D55E00}0.5 (0.1, 1.0)} & \cellcolor{Gray!10}{\color[HTML]{D55E00}0.6 (0.1, 1.0)} & \cellcolor{Gray!10}{\color[HTML]{D55E00}0.5 (0.1, 1.0)}\\
-1.1 (-3.3, 1.1) & -0.7 (-1.6, 0.2) & 0.1 & \color[HTML]{0072B2}-0.8 (-1.6, 0.0) & \color[HTML]{D55E00}-0.8 (-2.0, -0.0) & \color[HTML]{D55E00}-0.8 (-1.7, -0.1) & \color[HTML]{D55E00}-0.7 (-1.5, -0.0) & \color[HTML]{0072B2}-0.8 (-1.6, 0.0)\\
\cellcolor{Gray!10}{-0.3 (-0.7, 0.2)} & \cellcolor{Gray!10}{-0.3 (-0.7, 0.1)} & \cellcolor{Gray!10}{0.0} & \cellcolor{Gray!10}{\color[HTML]{0072B2}-0.3 (-0.6, 0.0)} & \cellcolor{Gray!10}{\color[HTML]{D55E00}-0.3 (-0.5, -0.0)} & \cellcolor{Gray!10}{\color[HTML]{D55E00}-0.3 (-0.5, -0.0)} & \cellcolor{Gray!10}{\color[HTML]{D55E00}-0.3 (-0.5, -0.0)} & \cellcolor{Gray!10}{\color[HTML]{0072B2}-0.3 (-0.6, 0.0)}\\
0.0 (-0.9, 0.9) & 0.1 (0.1, 0.1) & 0.0 & \color[HTML]{D55E00}0.1 (0.1, 0.1) & \color[HTML]{0072B2}0.1 (-0.3, 0.4) & \color[HTML]{D55E00}0.1 (0.1, 0.1) & \color[HTML]{D55E00}0.1 (0.1, 0.1) & \color[HTML]{D55E00}0.1 (0.1, 0.1)\\
\cellcolor{Gray!10}{0.7 (-1.7, 3.2)} & \cellcolor{Gray!10}{0.8 (0.5, 1.2)} & \cellcolor{Gray!10}{0.0} & \cellcolor{Gray!10}{\color[HTML]{D55E00}0.8 (0.5, 1.2)} & \cellcolor{Gray!10}{\color[HTML]{0072B2}0.8 (-0.2, 1.7)} & \cellcolor{Gray!10}{\color[HTML]{D55E00}0.8 (0.5, 1.2)} & \cellcolor{Gray!10}{\color[HTML]{D55E00}0.8 (0.5, 1.2)} & \cellcolor{Gray!10}{\color[HTML]{D55E00}0.8 (0.5, 1.2)}\\
-1.9 (-4.5, 0.8) & -2.0 (-5.3, 1.3) & 0.0 & \color[HTML]{0072B2}-1.9 (-4.0, 0.2) & \color[HTML]{D55E00}-1.9 (-3.8, -0.1) & \color[HTML]{D55E00}-1.9 (-3.7, -0.1) & \color[HTML]{D55E00}-1.9 (-3.7, -0.1) & \color[HTML]{0072B2}-1.9 (-4.0, 0.2)\\
\cellcolor{Gray!10}{0.5 (-0.3, 1.2)} & \cellcolor{Gray!10}{0.5 (-0.3, 1.3)} & \cellcolor{Gray!10}{0.0} & \cellcolor{Gray!10}{\color[HTML]{0072B2}0.5 (-0.1, 1.0)} & \cellcolor{Gray!10}{\color[HTML]{D55E00}0.5 (0.0, 1.0)} & \cellcolor{Gray!10}{\color[HTML]{D55E00}0.5 (0.0, 1.0)} & \cellcolor{Gray!10}{\color[HTML]{D55E00}0.5 (0.0, 1.0)} & \cellcolor{Gray!10}{\color[HTML]{0072B2}0.5 (-0.1, 1.0)}\\
0.4 (-0.3, 1.2) & 0.5 (-0.2, 1.3) & 0.0 & \color[HTML]{0072B2}0.5 (-0.0, 1.0) & \color[HTML]{D55E00}0.5 (0.0, 1.0) & \color[HTML]{D55E00}0.5 (0.0, 1.0) & \color[HTML]{D55E00}0.5 (0.0, 1.0) & \color[HTML]{0072B2}0.5 (-0.0, 1.0)\\
\cellcolor{Gray!10}{0.5 (-0.2, 1.3)} & \cellcolor{Gray!10}{0.5 (-0.2, 1.3)} & \cellcolor{Gray!10}{0.0} & \cellcolor{Gray!10}{\color[HTML]{0072B2}0.5 (-0.0, 1.1)} & \cellcolor{Gray!10}{\color[HTML]{D55E00}0.5 (0.1, 1.0)} & \cellcolor{Gray!10}{\color[HTML]{D55E00}0.5 (0.1, 1.0)} & \cellcolor{Gray!10}{\color[HTML]{D55E00}0.5 (0.1, 1.0)} & \cellcolor{Gray!10}{\color[HTML]{0072B2}0.5 (-0.0, 1.1)}\\
-0.5 (-1.3, 0.3) & -0.5 (-1.1, 0.2) & 0.0 & \color[HTML]{0072B2}-0.5 (-1.0, 0.0) & \color[HTML]{D55E00}-0.5 (-0.9, -0.1) & \color[HTML]{D55E00}-0.5 (-0.9, -0.1) & \color[HTML]{D55E00}-0.5 (-0.9, -0.1) & \color[HTML]{0072B2}-0.5 (-1.0, 0.0)\\
\cellcolor{Gray!10}{-0.5 (-0.9, 0.0)} & \cellcolor{Gray!10}{-0.3 (-1.2, 0.5)} & \cellcolor{Gray!10}{0.0} & \cellcolor{Gray!10}{\color[HTML]{D55E00}-0.4 (-0.8, -0.0)} & \cellcolor{Gray!10}{\color[HTML]{0072B2}-0.4 (-0.8, 0.0)} & \cellcolor{Gray!10}{\color[HTML]{D55E00}-0.4 (-0.8, -0.0)} & \cellcolor{Gray!10}{\color[HTML]{D55E00}-0.4 (-0.8, -0.0)} & \cellcolor{Gray!10}{\color[HTML]{D55E00}-0.4 (-0.8, -0.0)}\\
-0.3 (-5.0, 4.3) & -0.8 (-1.6, 0.0) & 0.0 & \color[HTML]{0072B2}-0.8 (-1.6, 0.0) & \color[HTML]{0072B2}-0.7 (-2.1, 1.5) & \color[HTML]{0072B2}-0.8 (-1.5, 0.1) & \color[HTML]{D55E00}-0.8 (-1.6, -0.0) & \color[HTML]{0072B2}-0.8 (-1.6, 0.0)\\
\cellcolor{Gray!10}{-0.7 (-1.9, 0.5)} & \cellcolor{Gray!10}{-0.7 (-1.7, 0.4)} & \cellcolor{Gray!10}{0.0} & \cellcolor{Gray!10}{\color[HTML]{0072B2}-0.7 (-1.5, 0.1)} & \cellcolor{Gray!10}{\color[HTML]{D55E00}-0.7 (-1.4, -0.0)} & \cellcolor{Gray!10}{\color[HTML]{D55E00}-0.7 (-1.4, -0.0)} & \cellcolor{Gray!10}{\color[HTML]{D55E00}-0.7 (-1.4, -0.0)} & \cellcolor{Gray!10}{\color[HTML]{0072B2}-0.7 (-1.5, 0.1)}\\
-6.0 (-12.5, 0.5) & -5.3 (-28.2, 17.6) & 0.0 & \color[HTML]{0072B2}-5.9 (-12.2, 0.3) & \color[HTML]{0072B2}-5.8 (-14.4, 3.6) & \color[HTML]{0072B2}-5.9 (-11.9, 0.3) & \color[HTML]{D55E00}-6.0 (-11.8, -0.1) & \color[HTML]{0072B2}-5.9 (-12.2, 0.3)\\*
\end{longtable}
\endgroup{}

\end{appendices}

\begin{spacing}{1}

\end{spacing}

\end{document}